\documentclass[preprint2]{aastex63}

\usepackage{xspace}
\usepackage{amssymb,amsmath}

\usepackage{longtable}
\usepackage{booktabs}

\newcommand{\GAIA}[0]{\textit{Gaia}\xspace}
\newcommand{\Ang}{\ensuremath{\text{\AA}}\xspace}

\shorttitle{YSO, Perseus, LAMOST, Gaia}
\shortauthors{X. L. Wang et al.}
\graphicspath{{./}{figures/}}

\begin{document}

\title{Distributed YSOs in the Perseus Molecular Cloud from the \textit{Gaia} and LAMOST Surveys}

\correspondingauthor{Min Fang}
\email{mfang@pmo.ac.cn}

\author{Xiao-Long Wang}
\affiliation{Purple Mountain Observatory, Chinese Academy of Sciences, No.10 Yuanhua Road, Qixia District, Nanjing 210023, People’s Republic of China; xlwang@pmo.ac.cn, 
mfang@pmo.ac.cn}
\affiliation{University of Chinese Academy of Sciences, No.19(A) Yuquan Road, Shijingshan District, Beijing 100049, People’s Republic of China}

\author{Min Fang}
\affiliation{Purple Mountain Observatory, Chinese Academy of Sciences, No.10 Yuanhua Road, Qixia District, Nanjing 210023, People’s Republic of China; xlwang@pmo.ac.cn, 
mfang@pmo.ac.cn}
\affiliation{School of Astronomy and Space Science, University of Science and Technology of China, Hefei, Anhui 230026, People’s Republic of China}

\author{Yu Gao}
\altaffiliation{Deceased}
\affiliation{Department of Astronomy, Xiamen University, Xiamen, Fujian 361005, People’s Republic of China}
\affiliation{Purple Mountain Observatory, Chinese Academy of Sciences, No.10 Yuanhua Road, Qixia District, Nanjing 210023, People’s Republic of China; xlwang@pmo.ac.cn, 
mfang@pmo.ac.cn}

\author{Hong-Xin Zhang}
\affiliation{Key Laboratory for Research in Galaxies and Cosmology, Department of Astronomy, University of Science and Technology of China, Hefei, Anhui 230026,People’s Republic of 
China}
\affiliation{School of Astronomy and Space Science, University of Science and Technology of China, Hefei, Anhui 230026, People’s Republic of China}

\author{Gregory J. Herczeg}
\affiliation{Department of Astronomy, Peking University, Yiheyuan Road 5, Haidian District, Beijing 100871, People’s Republic of China}
\affiliation{Kavli Institute for Astronomy and Astrophysics, Peking University, Yiheyuan Road 5, Haidian District, Beijing 100871, People’s Republic of China}

\author{Hong-Jun Ma}
\affiliation{Purple Mountain Observatory, Chinese Academy of Sciences, No.10 Yuanhua Road, Qixia District, Nanjing 210023, People’s Republic of China; xlwang@pmo.ac.cn, 
mfang@pmo.ac.cn}

\author{En Chen}
\affiliation{Purple Mountain Observatory, Chinese Academy of Sciences, No.10 Yuanhua Road, Qixia District, Nanjing 210023, People’s Republic of China; xlwang@pmo.ac.cn, 
mfang@pmo.ac.cn}
\affiliation{School of Astronomy and Space Science, University of Science and Technology of China, Hefei, Anhui 230026, People’s Republic of China}

\author{Xing-Yu Zhou}
\affiliation{Department of Astronomy, Peking University, Yiheyuan Road 5, Haidian District, Beijing 100871, People’s Republic of China}
\affiliation{Kavli Institute for Astronomy and Astrophysics, Peking University, Yiheyuan Road 5, Haidian District, Beijing 100871, People’s Republic of China}

\begin{abstract}

Identifying the young optically visible population in a star-forming region is essential for fully understanding the star formation event. In this paper, We identify 211 candidate members of the Perseus molecular cloud based on \GAIA astronomy. We use LAMOST spectra to confirm that 51 of these candidates are new members, bringing the total census of known members to 856. The newly confirmed members are less extincted than previously known members. Two new stellar aggregates are identified in our updated census. With the updated member list, we obtain a statistically significant distance gradient of $\rm 4.84\;pc\;deg^{-1}$ from west to east. Distances and extinction corrected color-magnitude diagrams indicate that NGC~1333 is significantly younger than IC~348 and the remaining cloud regions. The disk fraction in NGC~1333 is higher than elsewhere, consistent with its youngest age. The star formation scenario in the Perseus molecular cloud is investigated and the bulk motion of the distributed population is consistent with the cloud being swept away by the Per-Tau Shell.

\end{abstract}

\section{Introduction} \label{sec:intro}

A complete census of young stellar objects (YSOs) in star-forming regions is important for measuring the statistical properties of young stellar populations and for estimating the star formation rates \citep[e.g.,][]{Hsieh2013ApJS..205....5H,Young2015AJ....150...40Y,Mercimek2017AJ....153..214M}. Previous studies identified YSOs mainly based on their infrared (IR) excess emission \citep[e.g.,][]{Harvey2007ApJ...663.1149H,Evans2007C2D} or elevated X-ray emission \citep{Stelzer2012A&A...537A.135S}. Various color-color diagrams (CCDs) and color-magnitude diagrams (CMDs) are efficient in selecting sources with IR excess emission. However, these selections are always contaminated by non-YSO sources, such as broad-line active galactic nuclei (AGNs), star-forming galaxies, asymptotic giant branch (AGB) stars, that have similar IR properties as YSOs \citep{Robitaille2008AJ....136.2413R,Koenig2012ApJ...744..130K,Manara2018A&A...615L...1M,Herczeg2019ApJ...878..111H,Lee2021ApJ...916L..20L}. In addition, these CCDs and CMDs are inefficient in selecting evolved YSOs without circumstellar material, which have similar colors as main-sequence stars.

Accurate astrometric measurements from the \GAIA mission \citep{Gaia-Collaboration2016A&A...595A...1G} have changed this situation. With measurements of its distance and proper motion of individual sources, membership can be assessed directly, without any assumptions for the source properties. Different methods have been developed to identify new members in nearby star-forming regions, based on \GAIA astrometric data, including the following examples. Using the dataset from \GAIA DR2 \citep{Gaia-Collaboration2018A&A...616A...1G}, \citet{Canovas2019A&A...626A..80C} identified 166 new candidates in the $\rho$ Ophiuchi region applying different clustering algorithms, including \texttt{DBSCAN} \citep{Ester1996DBSCAN}, \texttt{HDBSCAN} \citep{Campello2013HDBSCAN,Campello2015HDBSCAN,McInnes2017JOSS....2..205M} and \texttt{OPTICS} \citep{Ankerst1999OPTICS}. Also in the $\rho$ Ophiuchi region, \citet{Grasser2021A&A...652A...2G} identified additional $\sim$200 new YSO candidates applying the \texttt{OCSVM} algorithm developed by \citet{Ratzenbock2020A&A...639A..64R}, based on the dataset from \GAIA EDR3 \citep{Gaia-Collaboration2021A&A...649A...1G}. Applying a simple thresholding method and utilizing the dataset from \GAIA DR2, \citet{Luhman2018AJ....156..271L} refined the sample of known members and identified new candidates in the Taurus star forming-region, \citet{Kubiak2021A&A...650A..48K} increased the census of known $\epsilon$ Cha members by more than 40\%. Both simple thresholding methods and clustering algorithms are powerful tools for identifying new members with high reliability in nearby star-forming regions.

The Perseus molecular cloud is one of the nearest ($\sim$300\;pc from the Sun, \citet{Ortiz-Leon2018ApJ...865...73O}) low mass star-forming regions \citep{Reipurth2008hsf1.book.....R}. It is a well-studied star-forming region \citep[e.g.,][]{Sargent1979ApJ...233..163S,Ladd1993ApJ...410..168L,Kirk2006ApJ...646.1009K,Ridge2006ApJ...643..932R,Enoch2007ApJ...666..982E,Arce2010ApJ...715.1170A} with a range of star-forming environments, both clustered (e.g., the two young clusters NGC~1333 and IC~348) and distributed, as well as several small dense clumps (B1, B5, L1448 and L1455) of the type that often produce one or a few stars \citep{Young2015AJ....150...40Y}. The Perseus molecular cloud has been observed in multi-wavelength studies (e.g., X-ray: 
\citealt{Winston2010AJ....140..266W,Alexander2011xru..conf..184A}; Optical: \citealt{Zhang2015RAA....15.1294Z}; IR: \citealt{Young2015AJ....150...40Y}; sub-millimeter: \citealt{Kirk2006ApJ...646.1009K,Enoch2007ApJ...666..982E}; radio: \citealt{Ridge2006ApJ...643..932R,Arce2010ApJ...715.1170A}). Many YSOs have been identified in this region previously \citep{Evans2009ApJS..181..321E,Hsieh2013ApJS..205....5H,Young2015AJ....150...40Y}. In a census of members of IC~348 and NGC~1333 using multi-epoch IRAC astrometry,
\citet{Luhman2016ApJ...827...52L} identified many members in the two clusters. However, for the remaining cloud regions, only IR excess emission was used to select large samples of YSO candidates (YSOc) \citep{Evans2009ApJS..181..321E,Hsieh2013ApJS..205....5H,Young2015AJ....150...40Y}, so many undiscovered YSOs without disks may be present in the remaining cloud regions. Recently, \citet{Pavlidou2021MNRAS.503.3232P} identified five new groups in the Perseus star-forming complex based on the astrometric data from \GAIA DR2 \citep{Gaia-Collaboration2018A&A...616A...1G}, but they mainly focused on the off-cloud regions specifically, instead of the main cloud regions outside the two young clusters.

In this paper, we use the \GAIA astrometry and photometry to identify candidate members in the Perseus molecular cloud, especially in the remaining cloud regions outside the two clusters, and use the LAMOST spectroscopic survey to confirm memberships of these candidates. We describe the datasets used in this work in Section~\ref{sec:data}, and the source selection procedures in Section~\ref{sec:source}. The results are presented in Section~\ref{sec:res} and discussed in Section~\ref{sec:discussion}. We give our summaries and conclusions in Section~\ref{sec:sum}.

\section{The Datasets}\label{sec:data}

\subsection{\GAIA Data}\label{sec:data:gaia}

We use the astrometric and photometric data from the \GAIA survey (\GAIA EDR3) \citep{Gaia-Collaboration2021A&A...649A...1G,Riello2021A&A...649A...3R, Fabricius2021A&A...649A...5F} to identify new candidates in the Perseus molecular cloud. We consider only the main cloud region that is covered by the $^{12}$CO J=1$\to$0 map from the COMPLETE (Coordinated Molecular Probe Line Extinction and Thermal Emission) Survey of Star-Forming Regions \citep{Ridge2006AJ....131.2921R}, corresponding to $157^{\circ}<l<161.2^{\circ}$ and $-22.2^{\circ}<b<-16.2^{\circ}$. \GAIA satellite provides us with very accurate astrometric measurements of almost two billion sources, including $\sim$60,000 within our survey area.

Based on the previously proposed distances to the Perseus molecular cloud, we include only sources with parallaxes between 2.2 and $5.0\ \rm mas$, corresponding to distances of about 200 to $450\ \rm pc$. Only sources whose observations are consistent with the five-parameter model \citep{Lindegren2018A&A...616A...2L} are retained for further analysis, that is we keep only sources with $ruwe<1.4$ \citep{Gaia-Collaboration2021A&A...649A...1G,Lindegren2018RUWE}. Additional quality cuts are applied to extract sources with high quality astrometry and photometry. Only sources with parallaxes over errors larger than 5 and with proper motion errors less than $0.5\;\rm mas$ are considered to produce astrometrically precise and reliable dataset. Sources with magnitude errors in $G$- or $RP$-bands larger than $0.1\;\rm mag$ are removed to produce high-quality catalog. With these cuts, we extract about 3000 \GAIA sources with high quality toward the direction of the Perseus molecular cloud.

\subsection{Optical Spectroscopic Data from The LAMOST Survey}\label{sec:data:lamost}

We use the spectroscopic data from the data release 7 of the LAMOST survey (LAMOST DR7) \citep{Luo2022yCat.5156....0L} to confirm memberships of the candidates in the cloud. LAMOST, the Large Sky Area Multi-Object Fiber Spectroscopic Telescope (also called the Guo Shoujing Telescope), is a quasi-meridian Schmidt telescope located at Xinglong Observatory Station in Hebei, China. The telescope has an effective aperture of 3.6$-$4.9\;m and a field of view of about $5^{\circ}$ in diameter. The telescope is equipped with 16 spectrographs and 4000 fibers, each spectrograph has a resolution of $\sim$2.5\;$\Ang$, and the wavelength coverage is 3690$-$9100\;$\Ang$ \citep{Cui2012RAA....12.1197C, Zhao2012RAA....12..723Z, Liu2015RAA....15.1089L}.

LAMOST DR7 contains more than 10 million spectra, more than 9.5 million of which are stellar spectra. In the direction of the Perseus molecular cloud, LAMOST obtained more than 9000 spectra of about 5500 distinct sources.

\subsection{Additional Ancillary Data}

The Perseus molecular cloud has been well covered by many large-sky surveys, including Pan-STARRS1 \citep{Hodapp2004AN....325..636H}, 2MASS \citep{Skrutskie2006AJ....131.1163S}, \textit{WISE} \citep{Wright2010AJ....140.1868W}. The Pan-STARRS1 (PS1) survey images the whole sky in five broadband filters, $grizy$, with a wavelength coverage from 0.4 to $1\;\mu\rm m$ \citep{Stubbs2010ApJS..191..376S}. We retrieve the $griz$ photometry from the PS1 DR1 catalog \citep{Chambers2016arXiv161205560C}. Sources that are brighter than $14\;\rm mag$\footnote{see \url{https://panstarrs.stsci.edu/}} or that have magnitude errors larger than $0.05\;\rm mag$, in $r$ or $i$ filters, are removed to avoid saturation problems and to produce high-quality catalog. The $JHK_{S}$ magnitudes are retrieved from the 2MASS All-Sky Point Source Catalog \citep{Skrutskie_2MASS_IPAC} which reaches limiting magnitudes of 15.8, 15.1 and $14.3\ \rm mag$ at 10$\sigma$ for $J$, $H$ and $K_{S}$-bands respectively \citep{Skrutskie2006AJ....131.1163S}. The \textit{WISE} survey is a mid-infrared full-sky survey undertaken in four bands: $W1$, $W2$, $W3$ and $W4$ bands with wavelengths centered at 3.35, 4.60, 11.56 and $22.09\ \micron$, respectively \citep{Wright2010AJ....140.1868W}. We take the \textit{WISE} photometry from the AllWISE source catalog \citep{Wright_AllWISE_IPAC} to determine the presence or absence of circumstellar disks around YSOs.

\section{Source Selection}\label{sec:source}

\subsection{Census from Previous Studies}\label{sec:known}

\begin{figure*}[!t]
    \centering
    \includegraphics[width=\textwidth]{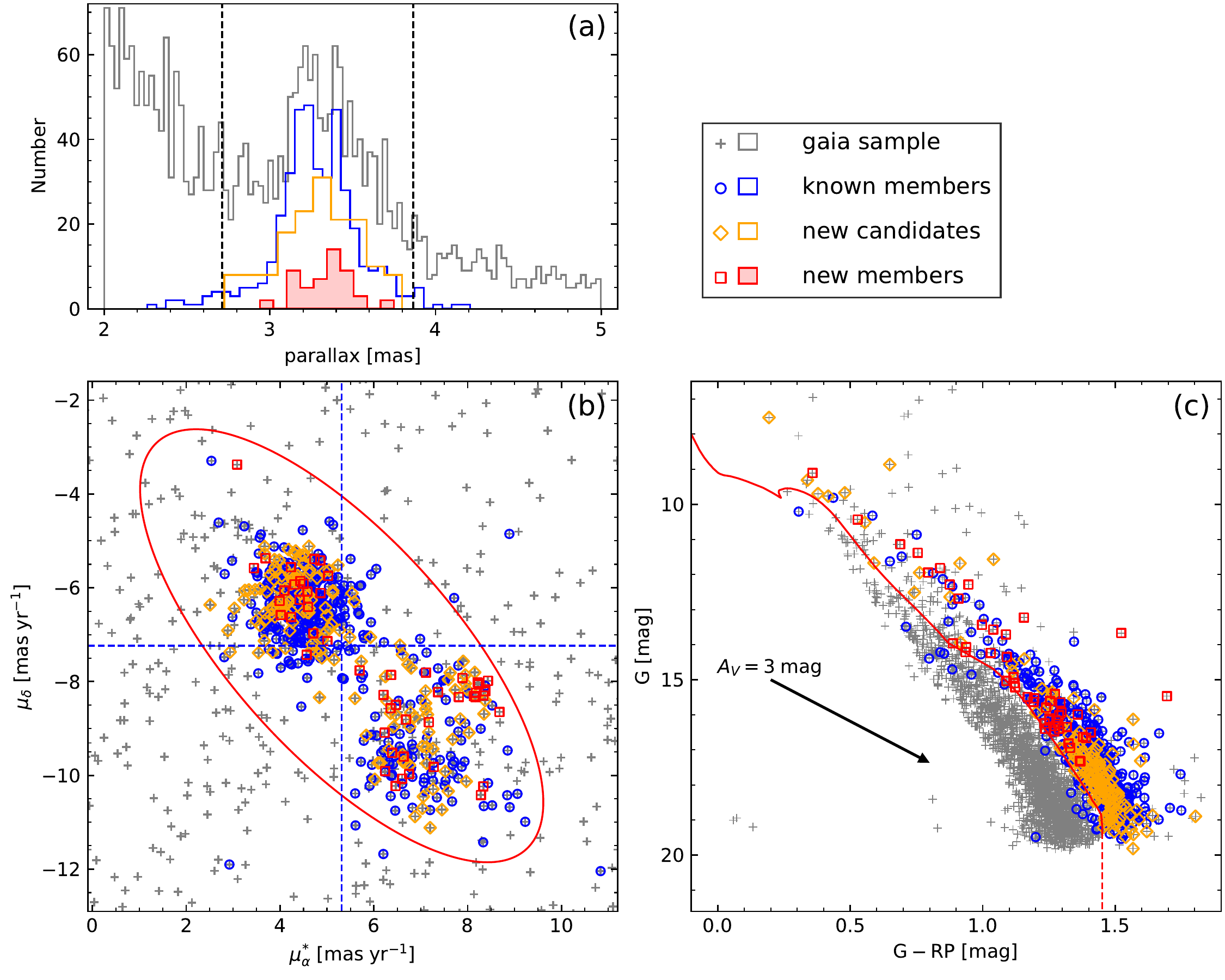}%
    \caption{The distributions of parallax (panel a), proper motions (panel b), as well as the $G-RP$ versus $G$ color-magnitude diagram (panel c) for the known members (blue circles and blue histogram), the new candidates (orange diamonds and orange histogram), the new members (red squares and red filled histogram), and the \GAIA sample described in Section~\ref{sec:data:gaia} (gray pluses and gray histogram). The vertical dashed lines in panel (a) mark the 3-$\sigma$ interval. The red ellipse in panel (b) is the 3-$\sigma$ confidence ellipse (see the text for detail). The blue vertical and horizontal dashed lines indicate the weighted means of the proper motions along right ascension and declination, respectively. The red solid line in panel (c) is the 10\;Myr isochrone from the PARSEC stellar model \citep{Bressan2012MNRAS.427..127B}, with solar metallicity and scaled to a distance of 300\;pc. The red dashed line corresponding to the color ($G-RP=1.45\;\rm mag$) of the lowest mass star in the PARSEC stellar model. The black arrow is the extinction vector of $A_{V}=3\;\rm mag$ (the typical extinction value of the Perseus sources, see Figure~\ref{fig:AVhist}) corresponding to the extinction law from \citet{Wang2019ApJ...877..116W}}
    \label{fig:plx_pm_ph}
\end{figure*}

\citet{Luhman2016ApJ...827...52L} compiled a thorough census for the two young clusters, and provided samples of 478 and 203 sources with well-confirmed memberships for IC~348 and NGC~1333, respectively. Beginning with this tabulation, we search the literature for additional members in the Perseus molecular cloud, especially in the regions outside the two clusters, with spectroscopically confirmed memberships. We add to our sample the 8 and 2 members identified by \citet{Esplin2017AJ....154..134E} and \citet{Luhman2020AJ....160...57L}, respectively. From the \citet{Kounkel2019AJ....157..196K} search for close companions around young stars, we include in our census the 114 members that have radial velocities between 0 and 25 $\rm km\;s^{-1}$. This velocity range is consistent with the radial velocities of members in IC~348 \citep{Cottaar2015ApJ...807...27C} and NGC~1333 \citep{Foster2015ApJ...799..136F}. \citet{Kounkel2019AJ....157..196K} reported radial velocity of $143\ \rm km\;s^{-1}$ for the source \texttt{2MASS J03453345+3145553}, but \citet{Cottaar2015ApJ...807...27C} reported radial velocity of around $20\ \rm km\;s^{-1}$ from multiple observations, so we retain this object in our member list. Combining all these tabulations together, we arrive at a sample of 805 known members with spectroscopically confirmed memberships in the Perseus molecular cloud.

Cross-matching this list of known members with the \GAIA catalog described in Section~\ref{sec:data:gaia}, using $2\arcsec$ tolerance, we obtain astrometric and photometric measurements for 406 known members ($\sim$50\% of the sample). It is not surprising that about half of the known members are not matched with the \GAIA catalog, since that we consider only sources with high-quality \GAIA measurements as described in Section~\ref{sec:data:gaia}. We would obtain 598 matches, i.e., $\sim$75\% of the sample, if no quality cuts were applied to the \GAIA catalog. In addition, many of the known members are either too faint ($\sim$150 members in our sample are fainter than $J=16\;\rm mag$, all without matches in the \GAIA catalog) or too embedded ($\sim$60 members in the census are deeply embedded class 0/I protostars) to be observed by the \GAIA satellite. 

Visually inspecting the properties of the 406 matched members, we find that the bulk of these members share common distances, motions and ages, with limited scatter, as expected from their memberships.  Figure~\ref{fig:plx_pm_ph} displays the distributions of parallax and proper motions, as well as the $G-RP$ versus $G$ color-magnitude diagram for these matched known members.

In the parallax distribution in Figure~\ref{fig:plx_pm_ph}, nearly all members have parallaxes within the 3-$\sigma$ interval from the weighted mean parallax.  The weighted mean and the weighted standard deviation of the parallax are $\overline{\varpi}=3.29\;\rm mas$ and $\sigma_{\varpi}=0.19\;\rm mas$, respectively, calculated by adopting the inverse of the parallax errors as the weight for each parallax.

In the proper motion distribution in Figure~\ref{fig:plx_pm_ph}, 394 of the matched members have proper motions within the 3-$\sigma$ confidence ellipse (the red ellipse in the plot). The confidence ellipse is calculated as follows. Visually inspecting the distribution of the proper motions, we identify six sources with extreme proper motions (outside the viewing range of the plot). In fact, these outliers are companions around young stars \citep{Kounkel2019AJ....157..196K}. The weighted means and the weighted standard deviations of the proper motion are $\overline{\mu_{\alpha}^{*}}=5.3\,\rm mas\,yr^{-1}$, $\overline{\mu_{\delta}}=-7.2\,\rm mas\,yr^{-1}$ and $\sigma_{\mu_{\alpha}^{*}}=1.43\,\rm mas\,yr^{-1}$, $\sigma_{\mu_{\delta}}=1.54\,\rm mas\,yr^{-1}$, respectively, calculated by adopting the inverse of the proper motion errors as the weight for each proper motion, excluding the six extreme cases. We also calculate the Pearson correlation coefficient of $-0.7$ between $\mu_{\alpha}^{*}$ and $\mu_{\delta}$ excluding the six extreme cases, considering measurement uncertainties of both axis, by using the Monte Carlo method proposed by \citet{Curran2014arXiv1411.3816C}. Using the weighted means, the weighted standard deviations and the correlation coefficient, we construct the 3-$\sigma$ confidence ellipse. 

From the CMD in Figure~\ref{fig:plx_pm_ph}, we find that all but one members are above or around the 10\;Myr isochrone from the PARSEC stellar model \citep{Bressan2012MNRAS.427..127B} with solar metallicity, scaled to a distance of 300\;pc. The only source well below the 10\;Myr isochrone is \texttt{2MASS J03291082+3116427}. Considering its location on the CMD and the presence of a disk \citep{Luhman2016ApJ...827...52L}, this object is likely harboring an edge-on disk.

\subsection{Candidates and New Members}\label{sec:can_and_new}

\begin{figure*}[!t]
    \centering
    \includegraphics[width=\textwidth]{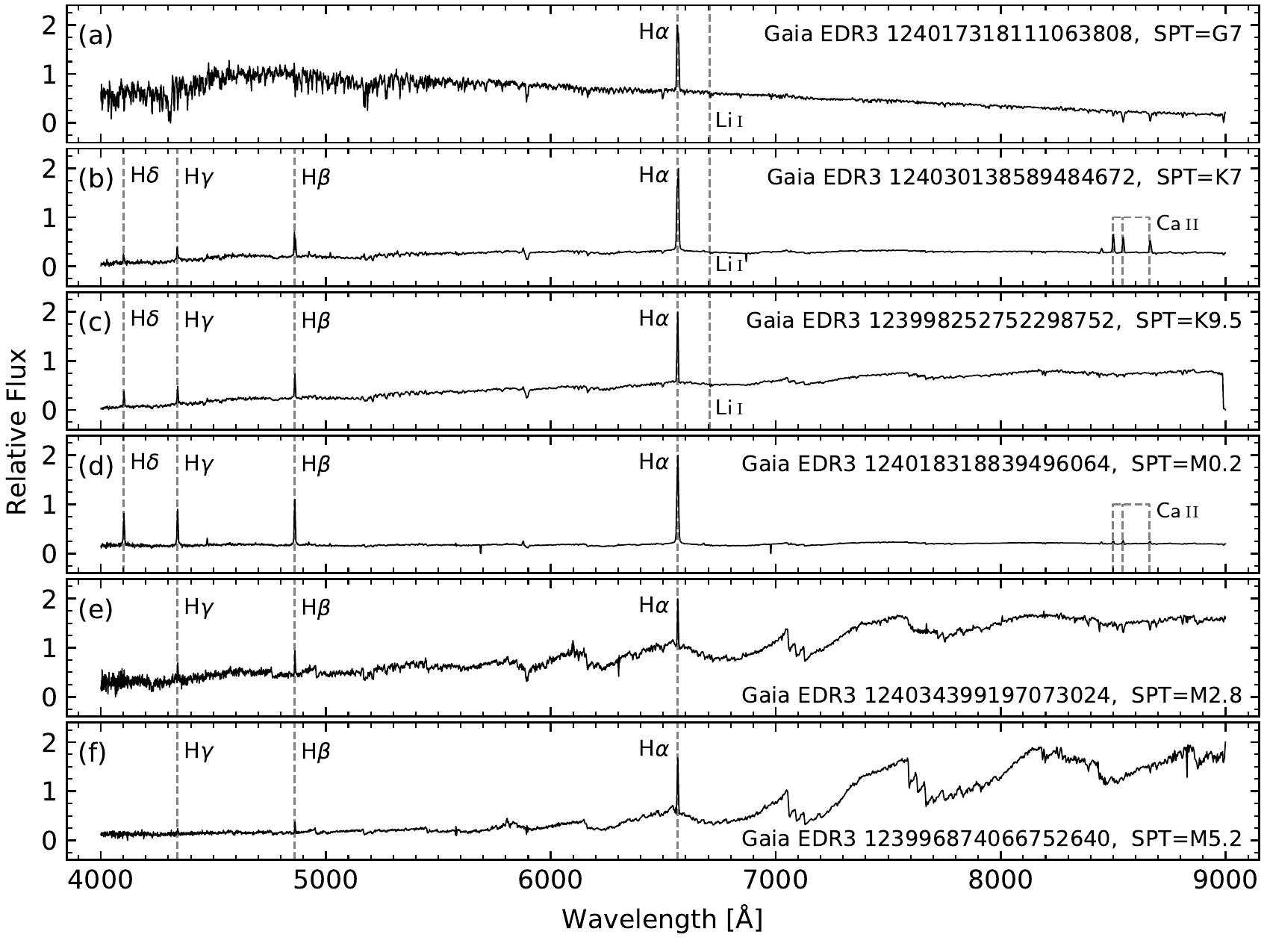}
    \caption{Example spectra of a representative sample of the newly confirmed members. The prominent emission lines and the Li\,\textsc{i}\,$\lambda\,6707\,\mathrm{\AA}$ absorption lines are marked in each panel, if present. The \GAIA EDR3 source designations and the corresponding spectral types are labeled as well.}
    \label{fig:SpecExam}
\end{figure*}

\begin{figure*}[!t]
    \centering
    \includegraphics[width=\textwidth]{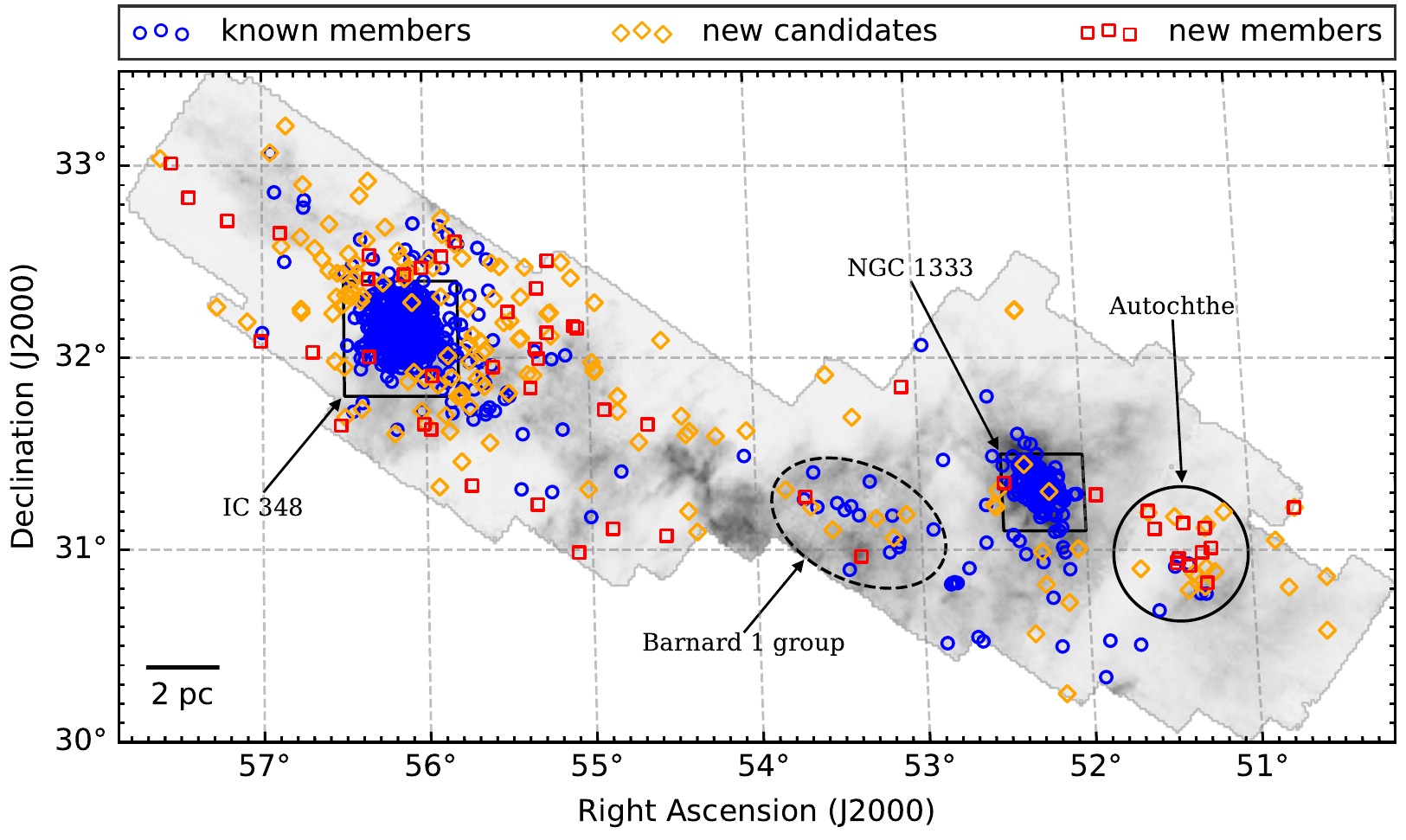}
    \caption{Spatial distribution of previously known members (blue circles), new members (red squares) and candidates (orange diamonds) in the Perseus molecular cloud overlaid on the FCRAO $^{12}$CO J=1$\to$0 integrated intensity map. The two rectangles mark the sky coverage of the two clusters as defined in \citet{Rebull2007ApJS..171..447R}, note that the cluster regions adopted here are slightly different from that in \citet{Luhman2016ApJ...827...52L}. The scale bar on the left bottom shows a size of $2\ \rm pc$ at a distance of $300\ \rm pc$. The large solid circle and the dashed ellipse mark the stellar aggregates discussed in Section~\ref{sec:clustering}.}
    \label{fig:spatial_distribution}
\end{figure*}

In the previous section, we find that nearly all the known members have parallaxes within the 3-$\sigma$ of the median parallax for the Perseus members and have proper motions within the 3-$\sigma$ confidence ellipse, as well as that all members are above or around the 10\;Myr isochrone. Based on these statistical properties of the known members, we identify new candidates from the \GAIA catalog. Specifically, we select a source as a new candidate if it has parallax between 2.71 and 3.87\;mas and its proper motion is within the 3-$\sigma$ confidence ellipse defined in the previous section. We also require that the candidate is above the 10\;Myr isochrone. As the PARSEC stellar model does not extend down to masses below the hydrogen burning limit, we artificially extend the 10\;Myr isochrone vertically to include low mass members. This extension line fits the low mass members fairly well. With these constraints, we select 211 additional candidate members from the \GAIA catalog. 

\citet{Kounkel2022arXiv220604703K} studied the Per OB2 association using \GAIA EDR3, applying the clustering algorithm \texttt{HDBSCAN} \citep{Campello2013HDBSCAN,Campello2015HDBSCAN,McInnes2017JOSS....2..205M}. In the same region as studied in our work, they discovered 129 new candidate members. In this work, we recover most of these candidates\footnote{Twenty-seven sources are excluded due to large $ruwe$ values, large photometric errors (Section~\ref{sec:data:gaia}) or being older than 10\;Myr}, and identified 109 new YSO candidates. These sources are distributed in the regions with low YSO surface density ($\sim50\;\mathrm{deg}^{-2}$), and are missed by the \texttt{HDBSCAN} algorithm used in \citet{Kounkel2022arXiv220604703K}.

By cross matching with the LAMOST archive, we obtain optical spectra for 55 candidates. Of these candidates, 38 have broad H$\alpha$ emission lines (characteristic of ongoing accretion activities \citep{Hartmann1994ApJ...426..669H,Muzerolle2001ApJ...550..944M}) or Li\,\textsc{i}\,$\lambda\,6707\,\mathrm{\AA}$ absorption lines (indicator of youth) in their optical spectra (as shown in Figure~\ref{fig:SpecExam}), indicating that they are bona fide members of the Perseus molecular cloud. Detailed analyses of these emission lines and corresponding accretion activities will be presented in a future paper (Wang, X.-L. et al., in preparation). For the remaining 17 candidates, we did not detect significant emission lines or Li\,\textsc{i}\,$\lambda\,6707\,\mathrm{\AA}$ absorption lines in their optical spectra. These 17 candidates include 8 A-type, 4 F-type, 2 G-type, 1 K-type and 2 M-type stars (see Section~\ref{sec:spt} for details of the spectral classification) and most of them show clear H$\alpha$ absorption lines in their optical spectra. Inspecting their locations on the Hertzsprung-Russell (H-R) diagram, we confirm 13 of them as young members.  The remaining four objects are classified as main-sequence (MS) stars or asymptotic giant branch (AGB) stars and are rejected from our list of candidates (see Section~\ref{sec:membership} for a discussion).

In summary, we identify 211 candidate members sharing common distances, motions and ages in the Perseus molecular cloud and confirm 51 of them as bona fide members, based on evidences of youth, including the presence of emission lines and Li\,\textsc{i}\,$\lambda\,6707\,\mathrm{\AA}$ absorption lines in their optical spectra, and their locations on the H-R diagram. We confirm more than 90\% (51/55) of the candidates having LAMOST spectra as Perseus members. These new members and candidates are listed in Table~\ref{tab:new}. The spatial distributions of all members (805 known members and 51 new members) and candidates (156 candidates) are displayed in Figure~\ref{fig:spatial_distribution}. Nearly all the members and candidates newly identified in this work are spread over the whole cloud outside the two young clusters.

\section{Stellar Properties}\label{sec:res}

\subsection{Spectral Types}\label{sec:spt}

\begin{figure}[!t]
    \centering
    \includegraphics[width=\columnwidth]{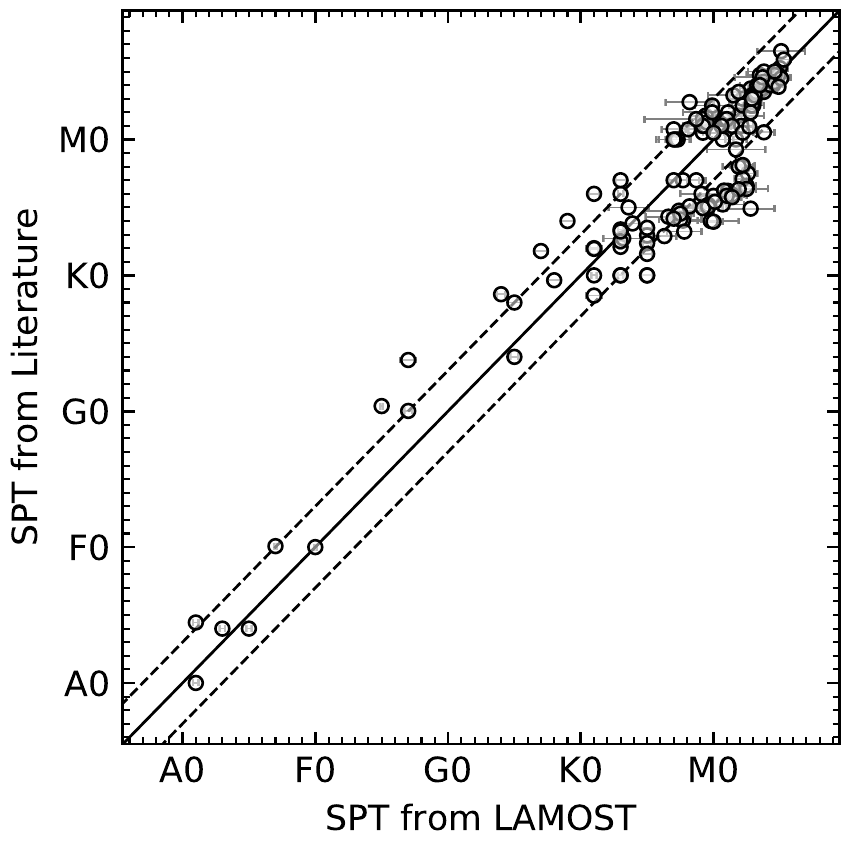}
    \caption{Comparison between spectral types from this work and that from previous works for known members that have LAMOST observations. The solid line indicates the line of equality, and the dashed lines mark differences of 3 sub-types.}
    \label{fig:SPTcom}
\end{figure}

In this section, we estimate spectral types for the newly confirmed members, listed in Table~\ref{tab:new}. Due to the complexity of classifying YSOs with late spectral types (emission lines and molecular bands), the LAMOST pipeline \citep{Wu2011A&A...525A..71W,Luo2015RAA....15.1095L} may give incorrect spectral types during the chi-square fitting procedures. Thus, for objects with automated spectral types from the LAMOST pipeline later than K0, we measured spectral type from the LAMOST spectra by applying the classification scheme from \citet{Fang2017AJ....153..188F}, which focuses on specific spectral features (i.e., VO, TiO and CaH-bands). For sources with earlier types, we adopt the spectral classifications from the LAMOST archive directly. To test the validity of these classification schemes, we apply these classification schemes to a sub-sample of the known members that have been observed by LAMOST as well. The spectral types obtained here and that from the literature are consistent with each other (Figure~\ref{fig:SPTcom}) .

\begin{figure}[!t]
    \centering
    \includegraphics[width=\columnwidth]{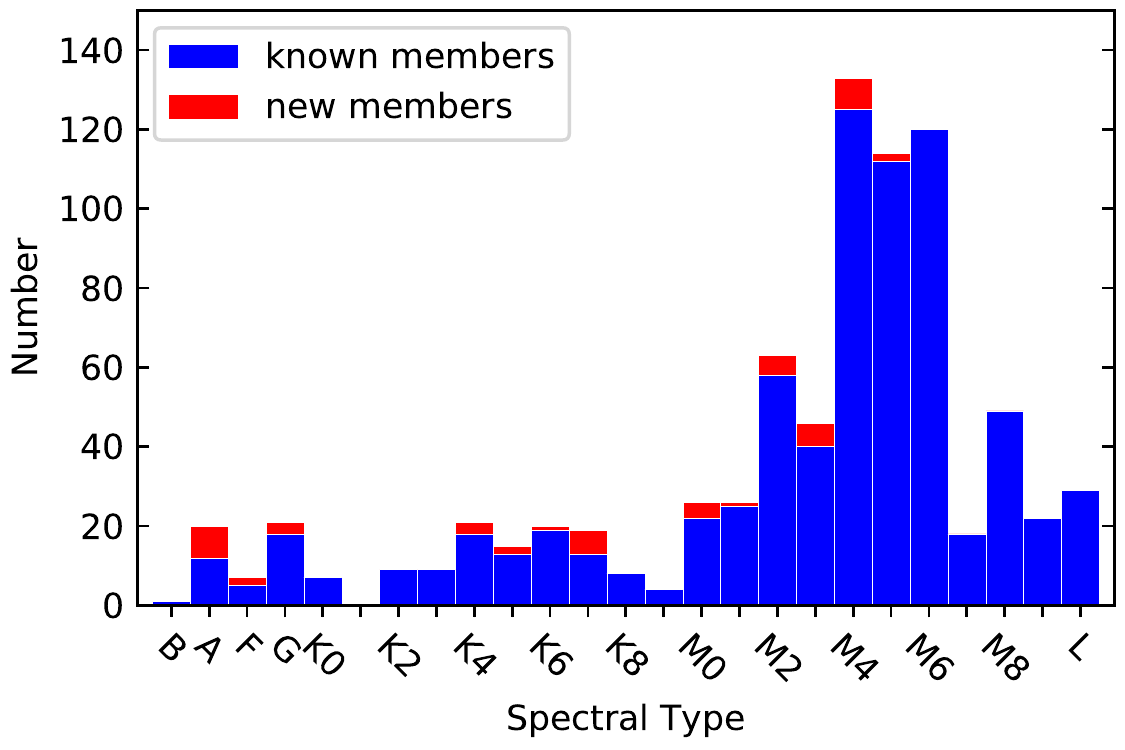}
    \caption{Distributions of spectral types for previously known (blue bars) and newly confirmed (red bars) members.}
    \label{fig:SPThist}
\end{figure}

The distributions of spectral types for previously known and newly confirmed members are displayed in Figure~\ref{fig:SPThist}. Most of the new members are late K and M types. Although we identify many A-, F- and G-type members, the overall distribution of the spectral types remains unchanged. Due to the sensitivity of the LAMOST survey, we did not select any new members later than about M6.

\subsection{Extinction Corrections}\label{sec:extinction}

The extinction of each star is determined individually. Intrinsic $g-r$, $r-i$, $i-z$, $z-J$, $J-H$, $H-K_{S}$ colors are estimated with corresponding spectral types for the stars with determined spectral types. For sources later than F0, the relationship from \citet{Fang2017AJ....153..188F} is adopted, and for earlier type stars that from \citet{Covey2007AJ....134.2398C} is used. These intrinsic colors are converted from the SDSS photometric system to the Pan-STARRS1 photometric system \citep{Tonry2012ApJ...750...99T}, and the extinction values are determined, combining with observed colors and the extinction law from \citet{Wang2019ApJ...877..116W}.

\begin{figure}[!t]
    \centering
    \includegraphics[width=\columnwidth]{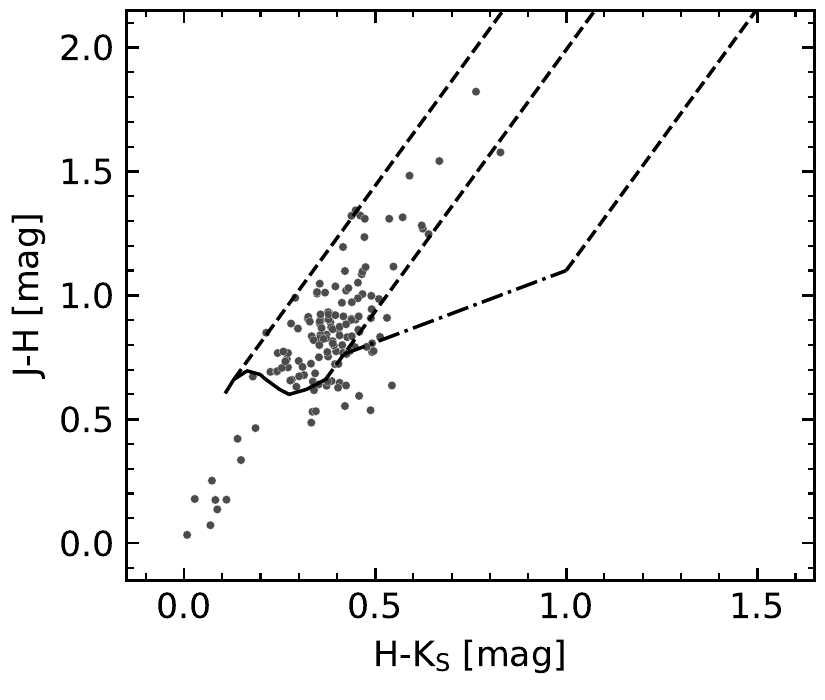}
    \caption{$H-K_{S}$ versus $J-H$ color-color diagram for the candidates in the Perseus molecular cloud. The solid line shows the intrinsic colors for main-sequence stars of K5-M6 types \citep{Bessell1988PASP..100.1134B}, and the dash-dotted line is the locus of T Tauri stars from \citet{Meyer1997AJ....114..288M}. The dashed lines are the reddening vectors from \citet{Wang2019ApJ...877..116W}.}
    \label{fig:CCD:HK_vs_JH}
\end{figure}

For sources lacking determined spectral types, we can only roughly determine their extinctions with the method described in \citet{Fang2013ApJS..207....5F}, in which the extinctions are obtained by employing the $H-K_{S}$ versus $J-H$ color-color diagram. Observed $J-H$ and $H-K_{S}$ colors are compared to intrinsic colors of main-sequence stars \citep{Bessell1988PASP..100.1134B} and T Tauri stars \citep{Meyer1997AJ....114..288M}, and extinction values are estimated for individual stars, using the extinction law from \citet{Wang2019ApJ...877..116W} (see Figure~\ref{fig:CCD:HK_vs_JH}). The typical error of the extinction values in $V$-band is $\sim$0.6\;mag and may be higher for stars with accretion disks.

\begin{figure}[!t]
    \centering
    \includegraphics[width=\columnwidth]{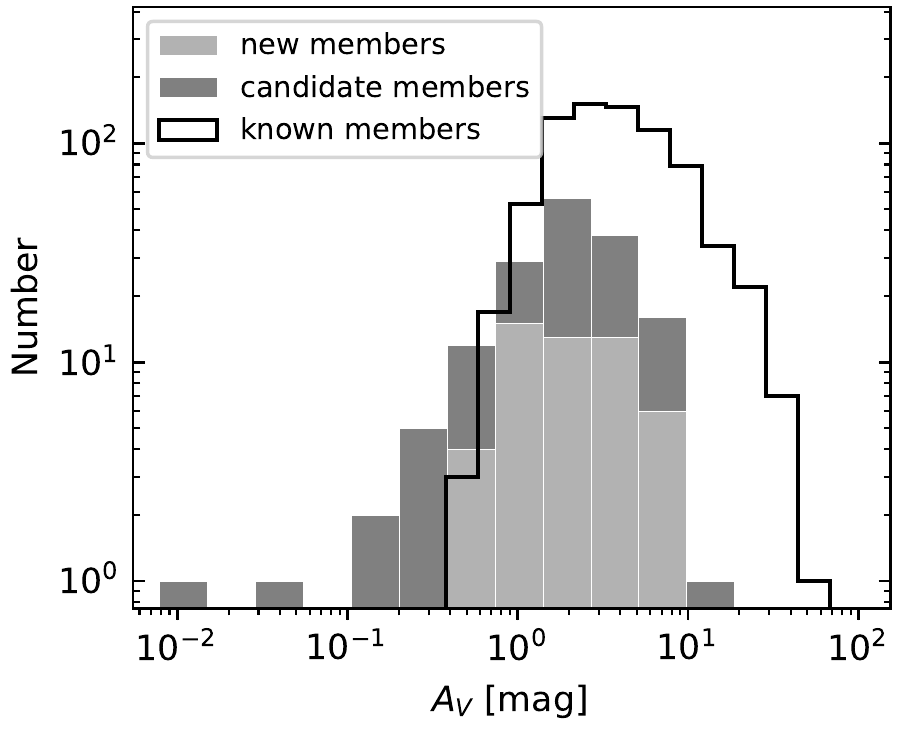}
    \caption{Histograms showing the extinction distributions for previously known members (open histogram), new members (light gray bars) and candidates (dark gray bars).}
    \label{fig:AVhist}
\end{figure}

These extinction values are listed in Table~\ref{tab:new} and displayed in Figure~\ref{fig:AVhist}. The median values of extinction for previously known members, new members and candidates are $A_{V}=3.6$, 2.2 and $2.2\;\rm mag$, respectively. A two sample KS-test also indicates that newly confirmed members and candidates have smaller extinction values than previously known sources. This result is consistent with the fact that most of our newly confirmed members and candidates are spread evenly across the cloud, located at less extincted regions, while most previous searches focused on the two clusters.

\section{Discussion}\label{sec:discussion}

\subsection{Membership Confirmations}\label{sec:membership}

We assess memberships for about 70\% of the candidates having LAMOST spectra as bona fide members based on the existence of H$\alpha$ emission lines or Li\,\textsc{i}\,$\lambda\,6707\,\mathrm{\AA}$ absorption lines in their optical spectra in Section~\ref{sec:can_and_new}. Similar as in \citet{Cieza2012ApJ...750..157C}, we assess memberships for the remaining candidates having LAMOST spectra, but lacking emission lines or Li\,\textsc{i}\,$\lambda\,6707\,\mathrm{\AA}$ absorption lines in their optical spectra, according to their locations on the H-R diagram in this section. To construct the H-R diagram, we convert the spectral types to effective temperatures using the empirical relations determined by \citet{Fang2017AJ....153..188F} for members later than F0-types and by \citet{Pecaut2013ApJS..208....9P} for earlier type members. The observed $J$ magnitudes are dereddened using the extinction values determined in Section~\ref{sec:extinction} and distance corrected assuming distances of 300\;pc for all sources (see the red filled histogram in panel (a) in Figure~\ref{fig:plx_pm_ph}). The stellar luminosities are then calculated as following.
\begin{equation}
\log\left(\dfrac{L_{\star}}{L_{\odot}}\right)=-\dfrac{BC_{J}+M_{J}-M_{\rm bol,\odot}}{2.5}
\end{equation}
where $M_{J}$ is the extinction corrected absolute $J$ magnitude, $BC_{J}$ is the bolometric correction in $J$ band, $M_{\rm bol,\odot}$ is the bolometric magnitude of the Sun. We take $M_{\rm bol,\odot}=4.755\;\rm mag$ \citep{Mamajek2012ApJ...754L..20M} in this work. The $BC_{J}$ values from \citet{Fang2017AJ....153..188F} are adopted for spectral types later than F0 and that from \citet{Pecaut2013ApJS..208....9P} for earlier spectral types.

\begin{figure}
    \centering
    \includegraphics[width=\columnwidth]{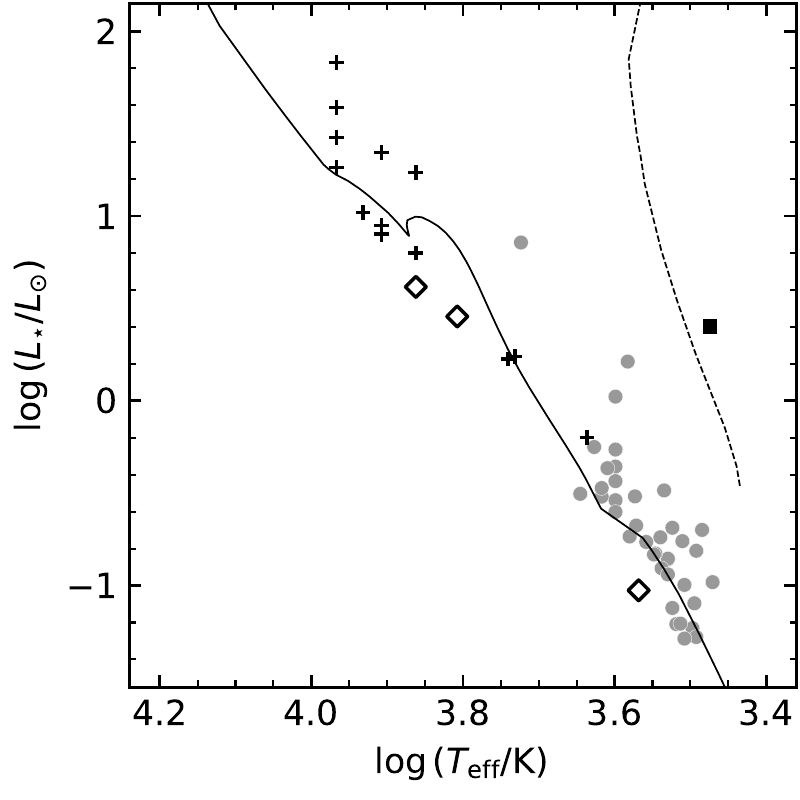}
    \caption{H-R diagram of members identified in this work. Members with and without H$\alpha$ emission lines or Li\,\textsc{i}\,$\lambda\,6707\,\mathrm{\AA}$ absorption lines in their optical spectra are marked with gray circles and black pluses, respectively. The solid line is the 10\;Myr isochrone and the dashed line is the stellar birth line, from the PARSEC stellar model \citep{Bressan2012MNRAS.427..127B}, with solar metallicity. The open diamonds and the square mark stars that are well below the 10\;Myr isochrone and that are well above the stellar birth line, respectively.}
    \label{fig:HRD_for_new}
\end{figure}

As shown in the H-R diagram (Figure~\ref{fig:HRD_for_new}), sources having emission lines or Li\,\textsc{i}\,$\lambda\,6707\,\mathrm{\AA}$ absorption lines in their optical spectra are all above or around the 10\;Myr isochrone, as expected from their memberships. Among the remaining 17 objects lacking emission lines or Li\,\textsc{i}\,$\lambda\,6707\,\mathrm{\AA}$ absorption lines in their optical spectra, 13 are located above or around the 10\;Myr isochrone and are consistent with being members in the Perseus molecular cloud, 3 are well below the 10\;Myr isochrone (marked with open diamonds), and are likely to be field main sequence stars, and one object is well above the stellar birth line (marked with square) and is a possible AGB star.

\subsection{Stellar Surface Densities and Substructures in the Perseus Cloud}\label{sec:clustering}

\begin{figure*}[!t]
    \centering
    \includegraphics[width=\textwidth]{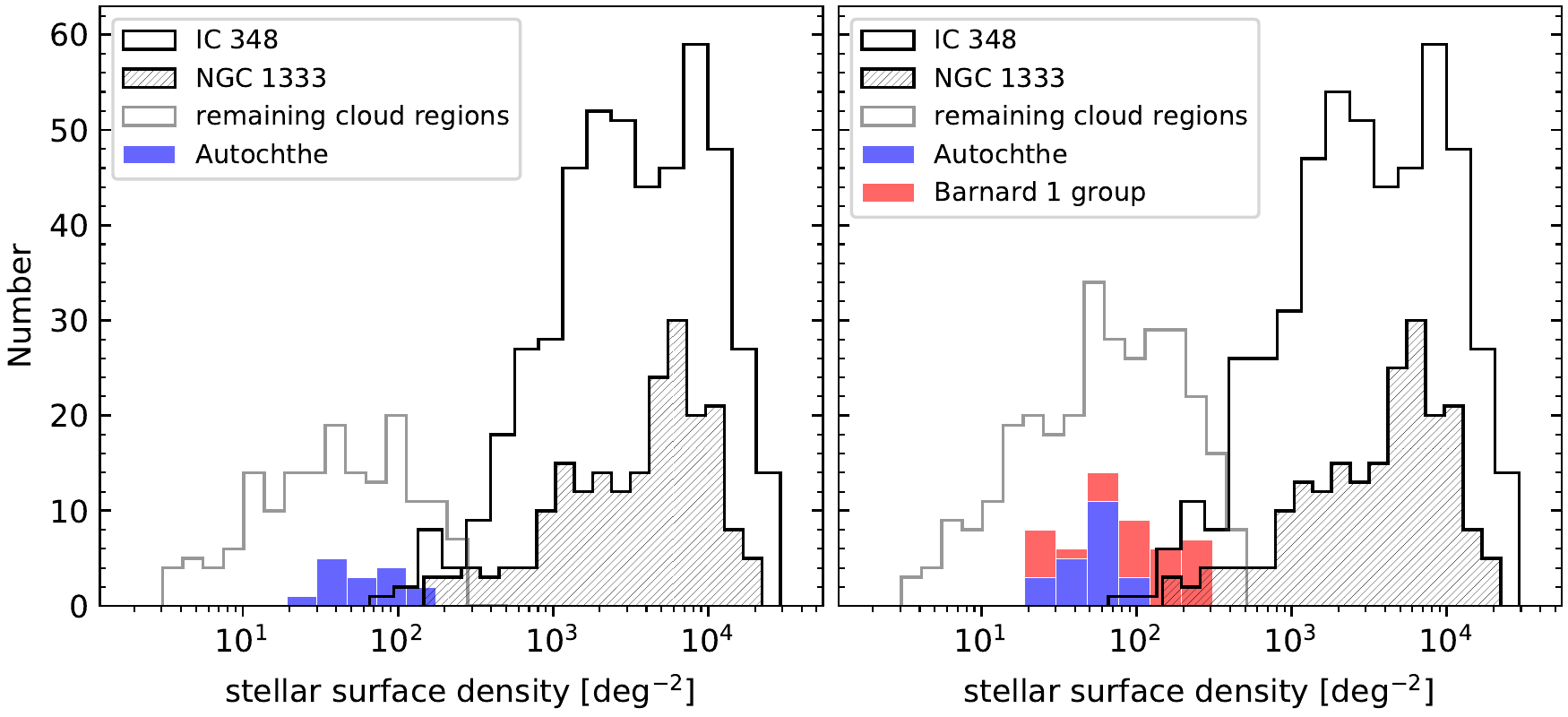}
    \caption{Left: histograms of the stellar surface densities for members in different cloud regions, i.e., IC~348 (black open histogram), NGC~1333 (hatched histogram), the remaining cloud regions (gray open histogram) and Autochthe (blue bars). Right: the same as in the left panel, but for densities including candidate members. The red bars indicate ``Barnard 1 group" found in Section~\ref{sec:clustering} and marked in Figure~\ref{fig:spatial_distribution}, stacked on top of that for Autochthe.}
    \label{fig:StellarDensity}
\end{figure*}

YSOs tend to form in clusters and are expected to have higher surface densities than field stars. As mentioned above, there are both clustered and distributed star formation in the Perseus molecular cloud. We have divided the cloud into three regions (i.e., NGC~1333, IC~348 and the remaining cloud regions, following \citealt{Rebull2007ApJS..171..447R}). We calculate the stellar surface densities $\Sigma$ for individual YSOs as \citep{Casertano1985ApJ...298...80C}, 
\begin{equation}
    \Sigma=\dfrac{n-1}{\pi r_{n}^{2}}
\end{equation}
where $n$ is the $n\rm th$ nearest star and $r_{n}$ is the distance to the $n\rm th$ nearest star. We adopt $n=6$ as a surface density reference, the same as that used in \citet{Gutermuth2009ApJS..184...18G}.

The resulting stellar surface density distributions are shown in Figure~\ref{fig:StellarDensity} for different cloud regions. The median densities are 3298, 4368 and $42\;\deg^{-2}$ in IC~348, NGC~1333 and the remaining cloud regions, respectively, if only confirmed members (previously known members and newly confirmed members) are included in the density calculations. The stellar surface densities are much higher in the two clusters than in the remaining cloud regions. The star formation environment in the remaining cloud regions is different from that in the clustered regions.  The remaining cloud regions represent an environment for distributed star formation. Including the candidate members in the calculation does not change the stellar surface densities in the cluster regions (3236 and $4378\;\deg^{-2}$ for IC~348 and NGC~1333 respectively), but nearly doubles the stellar density in the remaining cloud regions ($69\;\deg^{-2}$).

Visual inspection of Figure~\ref{fig:spatial_distribution} shows two additional groupings of YSO(c)s with enhanced stellar densities outside the main clusters IC~348 and NGC~1333. The stellar aggregate toward the western part of the cloud consists of 6 known members, 10 new members and 12 new candidates. This group is associated with the dark cloud L1448 \citep{Lynds1962ApJS....7....1L} and was originally identified by \citet{Pavlidou2021MNRAS.503.3232P}, who named the group Autochthe. The stellar aggregate to the east of NGC~1333 consists of 14 known members, 2 new members and 6 new candidates. This group is associated with the dark cloud Barnard 1 \citep{Barnard1919ApJ....49....1B} and we designate it as ``Barnard 1 group". These small groups represent the type of star formation that produce a few stars. The stellar densities for Autochthe and the ``Barnard 1 group'' are included in Figure~\ref{fig:StellarDensity}, these density distributions are consistent with the peaks in the distributions for the remaining cloud regions.

We also perform the density-based clustering algorithm \texttt{DBSCAN} \citep{Ester1996DBSCAN} on our dataset for further examination. When only confirmed members are considered, the algorithm returns 3 clusters, i.e., IC~348, NGC~1333 and Autochthe. The algorithm returned 4 clusters, i.e., IC~348, NGC~1333, Autochthe and ``Barnard 1 group", when the candidates identified in this work are included in the input dataset. These results are consistent with our visual inspection. 

\begin{figure}[!t]
    \centering
    \includegraphics[width=\columnwidth]{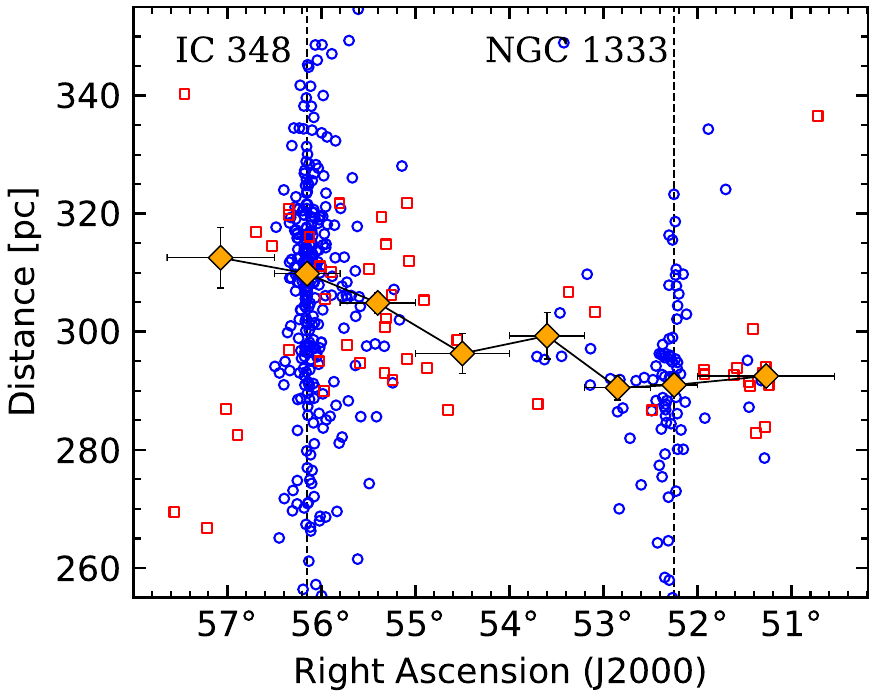}
    \caption{Distance changes along the right ascension direction. The known members and the new members are marked with blue circles and red squares, respectively. The two vertical dashed lines show the locations of the two clusters. The orange filled diamonds with error bars, connected with the solid line show the median distances in individual right ascension bins. The horizontal error bars are the corresponding right ascension ranges, the vertical error bars are the standard deviations of the median values in individual right ascension bins. The standard deviations are calculated from 10000 realizations assuming Gaussian errors \citep{Curran2014arXiv1411.3816C}.}
    \label{fig:RA_Plx}
\end{figure}

\subsection{Distance Gradient}\label{sec:dis_grad}

A single distance can not describe the whole cloud accurately, due to the complex structures of the Perseus molecular cloud. \citet{Ortiz-Leon2018ApJ...865...73O} obtained distances of 321\;pc and 293\;pc for IC~348 and NGC~1333 respectively, based on VLBA and \GAIA measurements.\footnote{Multiple measurements of the distance to the Perseus molecular cloud have been performed in the past. \citet{Cernis1990Ap&SS.166..315C} proposed a distance of 200$-$250\;pc to the Perseus molecular cloud according to extinction studies. Very long baseline interferometry studies of $\rm H_{2}O$ masers gave distances of $235\ \rm pc$ for NGC~1333 \citep{Hirota2008PASJ...60...37H} and $232\ \rm pc$ toward L1448 \citep{Hirota2011PASJ...63....1H}. A distance of 240\;pc was estimated by comparing the density of foreground stars with the \citet{Robin2003A&A...409..523R} Galactic model \citep{Lombardi2010A&A...512A..67L}. A distance of $250\ \rm pc$ toward the Perseus molecular cloud have been widely used by various studies \citep[e.g.,][]{Young2015AJ....150...40Y,Zhang2015RAA....15.1294Z,Enoch2006ApJ...638..293E}.}

In this work, we identify about 200 new members and candidates in the Perseus molecular cloud outside the two clusters, and these members spread nearly evenly across the cloud. With these distributed YSOs, as well as the clustered YSOs, we explore potential distance gradient toward the cloud. We measure distances of 310\;pc and 291\;pc toward IC~348 and NGC~1333, respectively, consistent with that measured by \citet{Ortiz-Leon2018ApJ...865...73O} within 1-sigma. In Figure~\ref{fig:RA_Plx}, we plot the changes of distances along the right ascension direction for all members. We also calculate the median distances of all members in corresponding right ascension bins. As shown in Figure~\ref{fig:RA_Plx}, the cloud is getting closer toward us gradually, from east to west. The Spearman's rank correlation coefficient between right ascension and distance is $|\rho|=0.88$ with $p=0.4\%$, indicating statistically significant correlation between right ascension and distance and that the distance gradient is real. The distance gradient is $4.84\;\rm pc\;deg^{-1}$ from NGC~1333 toward IC~348. 

\citet{Sargent1979ApJ...233..163S} found a smoothly varying LSR velocity across the cloud from their CO observations, and \citet{Enoch2006ApJ...638..293E} pointed out that there may also be a distance gradient across the cloud given the velocity gradient. However, we are unable to convert the LSR velocity gradient to a distance gradient directly, since the Perseus molecular cloud is located toward the Galactic anti-center. With our updated member list, we quantify this distance gradient. But it is worth noting that we can not rule out the possibility that the clusters reside in different clouds along the line of sight rather than a single cloud \citep{Bally2008hsf1.book..308B}.

\subsection{Evolutionary Stages in Different Cloud Regions}

\begin{figure*}[!t]
    \centering
    \includegraphics[width=\textwidth]{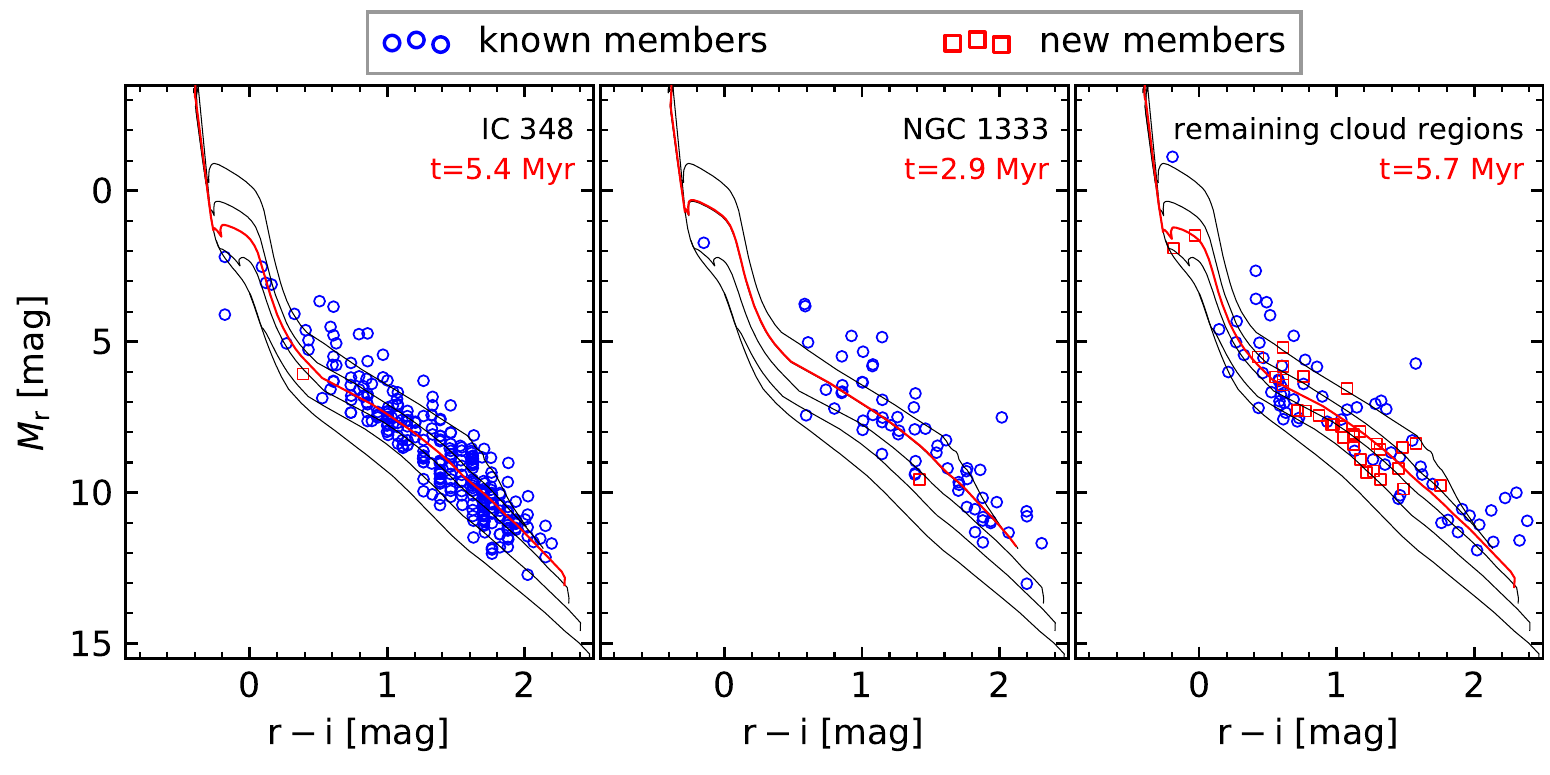}
    \caption{Pan-STARRS1 $r-i$ versus $r$ color-magnitude diagrams for known (blue circles) and new (red squares) members of the Perseus molecular cloud, divided into 3 regions, IC~348 (left), NGC~1333 (middle) and the remaining cloud regions (right). The points have been dereddened and distance corrected. In each panel, the black lines are 1, 3, 10, 30, 100 Myr isochrones, from top to bottom, from the PARSEC stellar model \citep{Bressan2012MNRAS.427..127B}, with solar metallicity, and the red solid line is the best fitting isochrone. The resulting ages are labeled.}
    \label{fig:CMDri}
\end{figure*}

\begin{figure*}[!t]
    \centering
    \includegraphics[width=\textwidth]{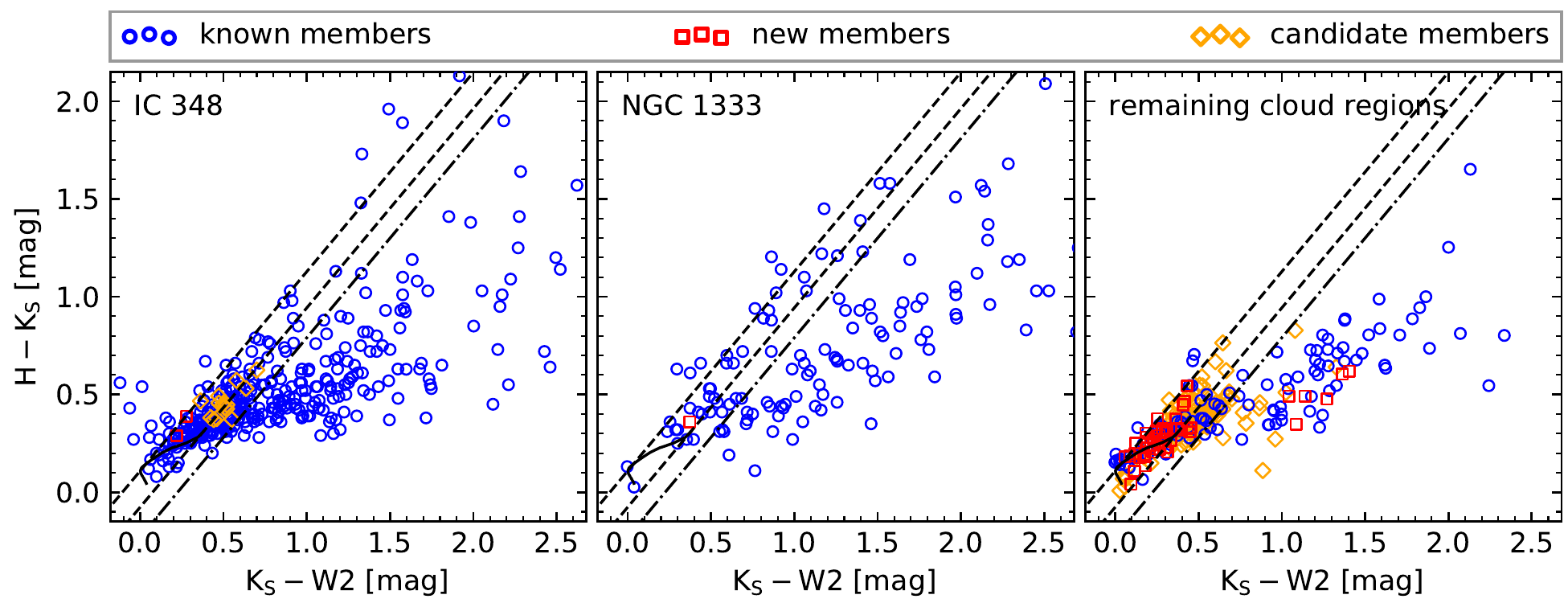}
    \caption{Left panel: infrared color-color plot for members and candidates in IC~348. The blue circles, red squares and orange diamonds mark known members, new members and candidate members, respectively. The black solid line is the locus of dwarfs from \citet{Pecaut2013ApJS..208....9P}, the dashed lines correspond to the extinction law from \citet{Wang2019ApJ...877..116W}, enclosing the color space of dwarfs due reddening. The dash-dotted line is 0.15\;mag (about 3 times the typical uncertainties in $K_{S}-W2$ colors) redder than the right border of the dwarf locus, serving as the boundary between disk and diskless YSOs. The middle and right panels are the same as the left panel, but for NGC~1333 and the remaining cloud regions, respectively.}
    \label{fig:CCD:disk}
\end{figure*}

In this section, we investigate the evolutionary stages in different cloud regions. \citet{Bell2013MNRAS.434..806B} measured an age of $\sim6\ \rm Myr$ for IC~348, whereas an age of $<1\ \rm Myr$ was measured for NGC~1333 \citep{Wilking2004AJ....127.1131W}. \citet{Young2015AJ....150...40Y} obtained similar conclusion based on the fractions of YSOs at different evolutionary stages. Dividing the Perseus molecular cloud into 3 regions, IC~348, NGC~1333 clusters and the remaining cloud regions \citep{Rebull2007ApJS..171..447R}, we construct the dereddened and distance corrected $r-i$ versus $r$ color-magnitude diagrams for each region (Figure~\ref{fig:CMDri}). For the cluster regions, we use the median distance for each cluster to do the distance correction, and individual distances are used for individual YSOs in the remaining cloud regions. We fit isochrones from the PARSEC stellar model \citep{Bressan2012MNRAS.427..127B} and estimate ages of 5.4, 2.9 and 5.7\;Myr for IC~348, NGC~1333 and the remaining cloud regions, respectively. The age for the remaining cloud regions may be underestimated, since we select only candidates above the 10\;Myr isochrone on the $G-RP$ versus $G$ color-magnitude diagram (Panel (c) in Figure~\ref{fig:plx_pm_ph}).

Recently, \citet{Kounkel2022arXiv220604703K} obtained much younger ages of 2.5\;Myr for both IC~348 and NGC~1333, by isochrone fitting of the \GAIA photometry to the MIST isochrones \citep{Choi2016ApJ...823..102C}. This discrepancy is mainly due to the difference between the PARSEC stellar model \citep{Bressan2012MNRAS.427..127B} and the MIST isochrones \citep{Choi2016ApJ...823..102C}. We obtain ages of 3.1 and 2.0\;Myr for IC~348 and NGC~1333 respectively performing isochrone fitting to the MIST isochrones \citep{Choi2016ApJ...823..102C}. We obtain much older age for IC~348 than for NGC~1333, which is consistent with a higher fraction of diskless objects in IC~348 than in NGC~1333 \citep{Young2015AJ....150...40Y,Luhman2016ApJ...827...52L}.

We obtain similar age for IC~348 as previous studies \citep{Bell2013MNRAS.434..806B, Luhman2016ApJ...827...52L}. \citet{Luhman2016ApJ...827...52L} proposed an older age for NGC~1333 than previous studies \citep[e.g.,][]{Wilking2004AJ....127.1131W}, and they attributed this discrepancy to the distance they assumed for NGC~1333. We obtain much younger age than \citet{Luhman2016ApJ...827...52L}, with our updated distance measurements. The remaining cloud regions are significantly older than NGC~1333 and nearly coeval with IC~348. \citet{Young2015AJ....150...40Y} found that the number ratio of PMS stars to protostars in the remaining cloud regions is similar to that in NGC~1333, but significantly smaller than that in IC~348. This discrepancy can be partly explained by the fact that we estimated ages based on optical CMDs, which may bias toward members at later evolutionary stages and thus older ages.

We estimate the fraction of sources harboring disks as another proxy for evolutionary stages of different cloud regions. \citet{Luhman2016ApJ...827...52L} estimated disk fractions of around 40\% and 60\% for IC~348 and NGC~1333 respectively, with disk fraction defined as $N(\mathrm{II})/N(\mathrm{II+III})$. Based on the source counts reported in \citet{Young2015AJ....150...40Y} (their Table~4), we estimate disk fractions of as high as 82\%, 92\% and 89\% for IC~348, NGC~1333 and the remaining cloud regions, respectively. The discrepancy between the two is mainly due to that \citet{Young2015AJ....150...40Y} used only infrared selected YSOs, whereas \citet{Luhman2016ApJ...827...52L} included many optically visible members, to calculate the disk fractions.

Since many of the new members and candidates identified in this work lack enough \textit{Spitzer} data to determine its presence or absence of a disk, we use the $K_{S}-W2$ versus $H-K_{S}$ color-color diagram (Figure~\ref{fig:CCD:disk}) to determine the presence or absence of a disk. As shown in the plot, sources between the two dashed lines are reddened dwarfs. We consider objects with $K_{S}-W2>0.98\times(H-K_{S})+0.22$ to be those harboring disks. We estimate disk fractions of 45\%, 65\% and 30\%$-$40\% for IC~348, NGC~1333 and the remaining cloud regions, respectively, with our updated member list. We estimate similar disk fractions as that estimated by \citet{Luhman2016ApJ...827...52L} for IC~348 and NGC~1333. Including or excluding the new members and candidates in the calculation has little impact on the disk fractions for IC~348 and NGC~1333, since we select few new members and candidates in these regions. But for the remaining cloud regions, we measure a much lower disk fraction compared to that excluding the new members and candidates in the calculation (30\%$-$40\% compared to 55\%). We find that the remaining cloud region is more coeval with IC~348 than with NGC~1333. This result suggests the importance of including \GAIA astrometric data in identifying YSOs at more evolved stages, with star-like spectral energy distributions.

\subsection{A Possible Scenario for Star Formation in the Perseus Molecular Cloud}\label{sec:SF}

\begin{figure*}[!t]
    \centering
    \includegraphics[width=\textwidth]{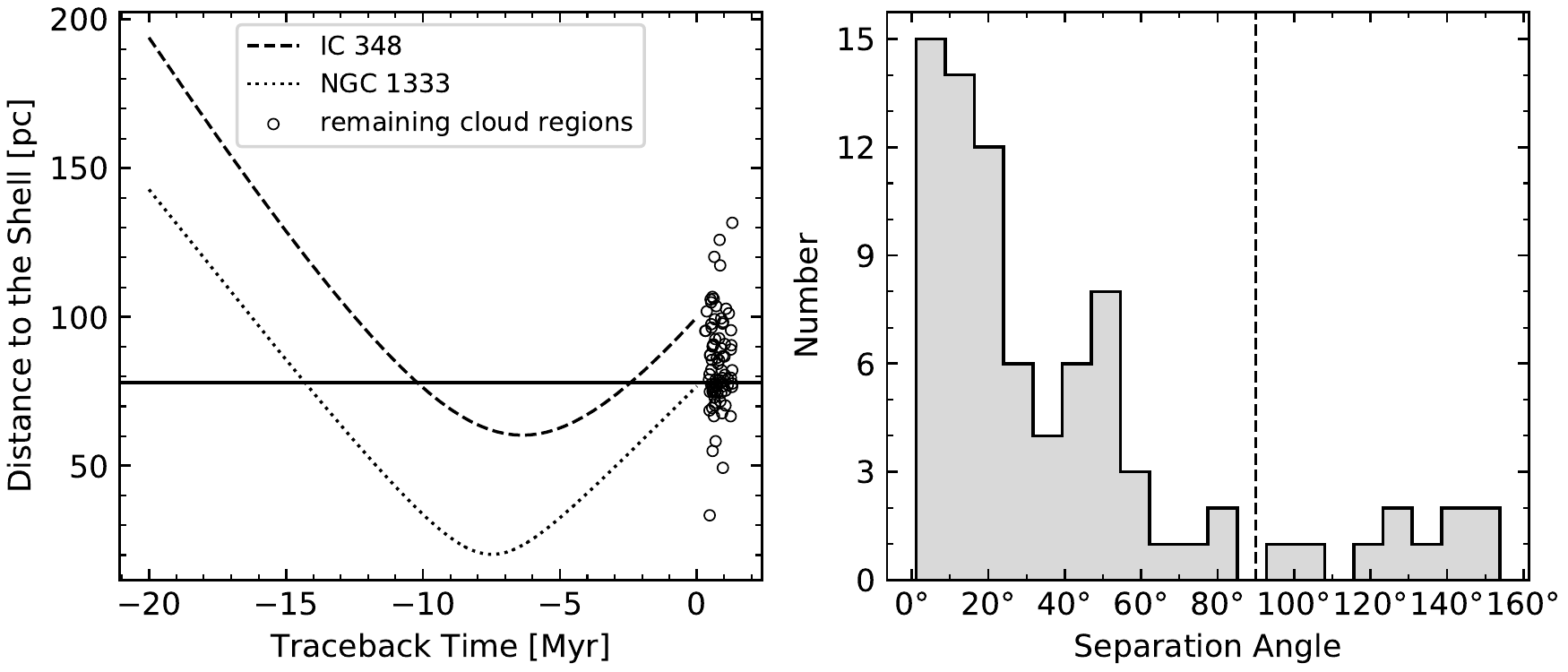}
    \caption{Left panel: distance to the center of the Per-Tau Shell as a function of traceback time. The dashed and dotted lines are stellar traceback lines from \citet{Zucker2022Natur.601..334Z} for IC~348 and NGC~1333 respectively. The horizontal solid line corresponds to a distance of 78\;pc, the radius of the shell \citep{Bialy2021ApJ...919L...5B}. The circles are individual members in the remaining cloud regions, shifted horizontally with random values to avoid overlap. Right panel: the distribution of separation angles between the velocity vectors and the radial vectors (pointing from the center of the shell to the star) of individual stars. The vertical dashed line indicates separation angle of $90^{\circ}$.}
    \label{fig:zucker}
\end{figure*}

\begin{figure}[!t]
    \centering
    \includegraphics[width=\columnwidth]{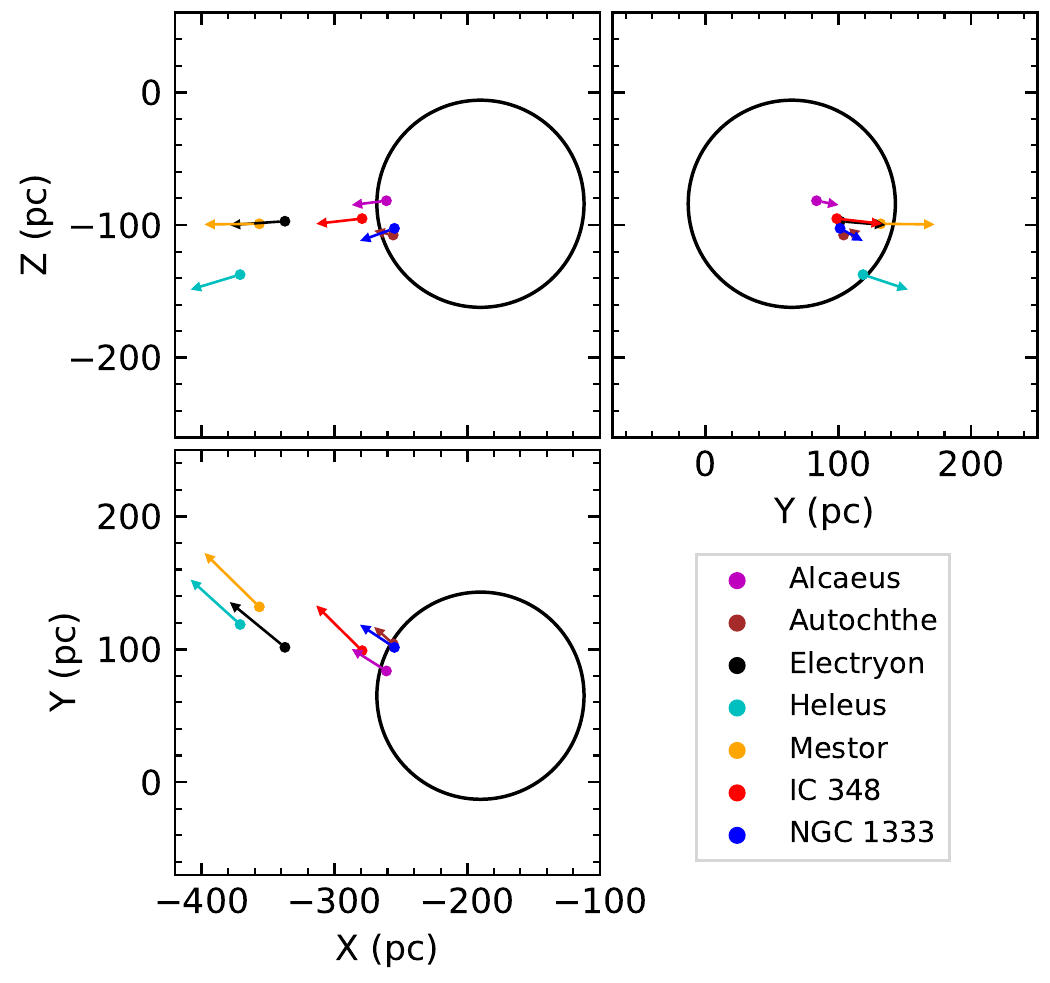}
    \caption{2D projections of 3D locations of the five groups identified by \citet{Pavlidou2021MNRAS.503.3232P} in a Heliocentric Galactic Cartesian reference frame. The two famous clusters IC~348 and NGC~1333 are also shown for comparison. Different colors corresponding to different regions as indicated in the legend. The arrows are the velocity vectors of individual groups or clusters. The large circle marks the Per-Tau Shell.}
    \label{fig:trace_for_P21}
\end{figure}

Recently, \citet{Zucker2022Natur.601..334Z} proposed that \textit{star formation near the Sun is driven by expansion of the Local Bubble} \citep{Cox1987ARA&A..25..303C} with stellar traceback method. They found that nearly every well-known molecular cloud in the vicinity of the Sun lies on the surface of the Local Bubble, except the Perseus molecular cloud. They attributed this exception to the displacement of the recently discovered Per-Tau Shell \citep{Bialy2021ApJ...919L...5B}, containing Perseus on its far side and Taurus on its near side. The stellar traceback data provided by \citet{Zucker2022Natur.601..334Z} indicate that IC~348 and NGC~1333 are moving away from the center of the Per-Tau Shell since about 6$-$8\;Myr ago, and that IC~348 has already escaped the shell whereas NGC~1333 is passing through the surface of the shell (as illustrated in the left panel of Figure~\ref{fig:zucker}).

Since we double the census of members in the remaining cloud regions, we can investigate the distributed populations with respect to the feedback from the Per-Tau Shell or the Local Bubble. As these distributed populations spread over large areas, it is improper to apply the stellar traceback method to these data, considering the remaining cloud regions as a whole. We investigate motions and locations of individual members relative to the center of the Per-Tau Shell, nearly all of these distributed members have velocities pointing outward the shell with separation angles less than $90^{\circ}$ (as shown in the right panel of Figure~\ref{fig:zucker}), about 60\% of them have escaped the shell (see the scatter plot in the left panel of Figure~\ref{fig:zucker}). These facts may point to the scenario that expansion of the Local Bubble drives the formation of the surface clouds \citep{Zucker2022Natur.601..334Z} as well as the Perseus molecular cloud, and subsequent expansion of the Per-Tau Shell blows away the Perseus molecular cloud from the Local Bubble's surface and results in elongated shape of the cloud. This elongation may partly explain the distance gradient confirmed in Section~\ref{sec:dis_grad}.

We have searched the literature as well as the \GAIA archive for possible massive stars in the vicinity of the cloud that may be responsible for creating the Per-Tau Shell, but no massive stars were found within the region we studied. \citet{Bialy2021ApJ...919L...5B} estimated that 2-8 supernova may be responsible for creating such a large shell structure. Since our study is focused on the main cloud region only, detailed analysis of searching for the driving source of the shell is beyond the scope of current study.

Visualizing the five groups identified by \citet{Pavlidou2021MNRAS.503.3232P} (see Figure~\ref{fig:trace_for_P21}), we find that the three groups, Electryon, Heleus and Mestor are far away from the Per-Tau Shell, the remaining 2 groups Alcaeus and Autochthe, as well as NGC~1333 and IC~348 are near the surface of the shell. Considering their ages (3-5\;Myr, \citet{Pavlidou2021MNRAS.503.3232P}) and their distances to the shell, the formation of the three off-shell groups are irrelevant to the Per-Tau Shell's expansion. Alcaeus share similar age as IC~348 and Autochthe is coeval with NGC~1333. These groups may form during the same star formation event.

\section{SUMMARY}\label{sec:sum}

We have performed a search for new members and candidates in the Perseus molecular cloud, using spectroscopic data from the LAMOST survey and astrometric data from the \GAIA survey. The spectral types and extinction corrections are determined for the newly confirmed members. The spatial distributions of the updated member list and the age differences in different cloud regions are investigated. The main contributions are summarized as follows.

\begin{enumerate}
    \item We have performed a thorough census for new members and candidates in the Perseus molecular cloud and identified 211 candidate members sharing common distances, motions and ages. We identify 51 candidates as new members of the cloud based on evidence of youth including the presence of emission lines or Li\,\textsc{i}\,$\lambda\,6707\,\mathrm{\AA}$ absorption lines in their optical spectra and their locations on the H-R diagram, which brings the total number of known members to 856. Due to the sensitivity of the LAMOST survey, we don't identify any new members later than M6.
    \item The stellar aggregate named Autochthe in \citet{Pavlidou2021MNRAS.503.3232P} is confirmed as a real coeval stellar aggregate with a dozen of confirmed members, with our revised member list. We also identify another small aggregate associated with the dark cloud Barnard 1.
    \item The new members are less extincted, and reside in regions with low surface densities. Cloud members in the remaining cloud regions represent the type of distributed star formation, as well as star formation of the type that often produce one or a few stars.
    \item A statistically significant distance gradient of $\rm 4.84\;pc\;deg^{-1}$ is measured from west to east. But it is worth noting that the possibility that the two clusters reside in different clouds along the line of sight rather than a single cloud still exists.
    \item We estimate similar ages for IC~348 and the remaining cloud regions and NGC~1333 is much younger than the two. The disk fraction in NGC~1333 is higher than in elsewhere, consistent with its youngest age.
    \item The bulk motion of the distributed populations is pointed outward of the Per-Tau Shell, and the five groups identified by \citet{Pavlidou2021MNRAS.503.3232P} are moving away from the shell as well. These are consistent with the star formation scenario proposed by \citet{Zucker2022Natur.601..334Z}.
\end{enumerate}

\acknowledgements
We acknowledge the support of the National Natural Science Foundation of China (NSFC, Grant No. 11390373) and National Key R\&D Program of China (No. 2017YFA0402700). HZ thanks the support from the FONDECYT Postdoctoral Grant (No. 3160538). GJH acknowledges support from support of the National Natural Science Foundation of China (NSFC, Grant No. 12173003). We thank A-Li Luo, Jian-Jun Chen, Wen Hou and other staffs from NAOC for their help of downloading the data and comments on the data reduction. This work has made use of data from the European Space Agency (ESA) mission {\it Gaia} (\url{https://www.cosmos.esa.int/gaia}), processed by the \GAIA Data Processing and Analysis Consortium (DPAC, \url{https://www.cosmos.esa.int/web/gaia/dpac/consortium}). Funding for the DPAC has been provided by national institutions, in particular the institutions participating in the \GAIA Multilateral Agreement. Guoshoujing Telescope (the Large Sky Area Multi-Object Fiber Spectroscopic Telescope LAMOST) is a National Major Scientific Project built by the Chinese Academy of Sciences. Funding for the project has been provided by the National Development and Reform Commission. LAMOST is operated and managed by the National Astronomical Observatories, Chinese Academy of Sciences. This research made use of \texttt{APLpy}, an open-source plotting package for Python \citep{aplpy2012}. This research made use of \texttt{Astropy}, a community-developed core Python package for Astronomy \citep{Astropy-Collaboration2013A&A...558A..33A,Astropy-Collaboration2018AJ....156..123A}. We also acknowledge the various Python packages that were used in the data analysis of this work, including \texttt{Matplotlib} \citep{matplotlib2007}, \texttt{NumPy} \citep{numpy2020}, \texttt{SciPy} \citep{scipy2020}, \texttt{Shapely} \citep{shapely2007}, \texttt{Scikit-Learn} \citep{Pedregosa2011sklearn}.



\setlength{\tabcolsep}{0.8ex}\renewcommand{\arraystretch}{1.0}
\footnotesize
\begin{longtable*}{ccccccccc}
\caption[]{Properties of new members and candidates}\label{tab:new}\\\toprule
Gaia EDR3 & $\alpha$ & $\delta$ & $\varpi$ & $\mu_{\alpha}^{*}$                     & $\mu_{\delta}$                         & SPT & $A_{V}$ & membership \\[1ex]
          & ($\deg$) & ($\deg$) & (mas)    & $\left(\dfrac{\rm mas}{\rm yr}\right)$ & $\left(\dfrac{\rm mas}{\rm yr}\right)$ &     & (mag)   &             \\
\midrule
\endfirsthead
\caption[]{--\,{\it continued}}\\\toprule
Gaia EDR3 & $\alpha$ & $\delta$ & $\varpi$ & $\mu_{\alpha}^{*}$                     & $\mu_{\delta}$                         & SPT & $A_{V}$ & membership \\[1ex]
          & ($\deg$) & ($\deg$) & (mas)    & $\left(\dfrac{\rm mas}{\rm yr}\right)$ & $\left(\dfrac{\rm mas}{\rm yr}\right)$ &     & (mag)   &             \\
\midrule
\endhead
\bottomrule
\endfoot
\endlastfoot
120447379950667008 & 55.085941 & 30.986461 & 3.39 & 6.37 & -7.86  & $\rm K9.7\pm1.2$ & 0.91     & member    \\
120460024334304128 & 54.553027 & 31.074960 & 3.35 & 6.20 & -8.28  & $\rm M3.7\pm0.5$ & 1.04     & member    \\
120463116710737920 & 54.875402 & 31.110591 & 3.40 & 6.69 & -8.81  & $\rm M4.2\pm0.5$ & 0.89     & member    \\
120903501181107968 & 52.162684 & 30.251784 & 3.45 & 7.97 & -10.13 & $\cdots$         & $\cdots$ & candidate \\
120934910278041600 & 52.333973 & 30.563745 & 3.53 & 7.59 & -10.34 & $\cdots$         & 0.42     & candidate \\
120952429448891520 & 52.122005 & 30.726234 & 3.05 & 4.18 & -5.18  & $\cdots$         & 3.45     & candidate \\
120959816792685312 & 52.255915 & 30.820725 & 3.42 & 7.05 & -10.45 & $\cdots$         & $\cdots$ & candidate \\
120996959670049152 & 51.678603 & 30.902172 & 3.29 & 7.03 & -10.70 & $\cdots$         & 0.38     & candidate \\
121007546764868864 & 52.051515 & 31.006618 & 3.46 & 5.93 & -7.06  & $\cdots$         & $\cdots$ & candidate \\
121007546764868992 & 52.050552 & 31.006813 & 3.51 & 6.56 & -7.22  & $\cdots$         & $\cdots$ & candidate \\
121010329903672576 & 52.272779 & 30.990384 & 3.42 & 6.61 & -7.43  & $\cdots$         & 2.90     & candidate \\
121147871936317184 & 53.372048 & 30.966361 & 3.26 & 7.11 & -7.81  & $\rm K4\pm0.4$   & 4.92     & member    \\
121155087481238656 & 53.540261 & 31.104290 & 3.28 & 7.11 & -8.59  & $\cdots$         & 2.33     & candidate \\
121163501321772672 & 53.170418 & 31.063100 & 3.52 & 7.26 & -10.29 & $\cdots$         & 1.22     & candidate \\
121165837783918848 & 53.273082 & 31.164778 & 3.69 & 8.16 & -8.30  & $\cdots$         & 3.22     & candidate \\
121190920393107840 & 54.362957 & 31.093630 & 3.53 & 6.47 & -8.71  & $\cdots$         & 0.69     & candidate \\
121215079583836800 & 54.412848 & 31.201997 & 3.69 & 5.84 & -9.15  & $\cdots$         & $\cdots$ & candidate \\
121242807893626368 & 54.416757 & 31.599503 & 3.55 & 7.14 & -8.70  & $\cdots$         & 0.26     & candidate \\
121243014052056320 & 54.395363 & 31.618091 & 3.58 & 6.43 & -8.96  & $\cdots$         & $\cdots$ & candidate \\
121251049934719616 & 53.667472 & 31.223274 & 3.48 & 7.10 & -8.46  & $\cdots$         & $\cdots$ & candidate \\
121251810145061376 & 53.700758 & 31.276131 & 3.47 & 7.37 & -8.22  & $\rm M4.3\pm0.9$ & 0.94     & member    \\
121252875296947584 & 53.818389 & 31.312865 & 3.64 & 6.36 & -8.68  & $\cdots$         & 0.26     & candidate \\
121284481960254976 & 54.235610 & 31.592405 & 3.30 & 3.47 & -6.04  & $\cdots$         & 2.33     & candidate \\
121286994517121792 & 54.047228 & 31.620840 & 3.34 & 5.60 & -8.36  & $\cdots$         & $\cdots$ & candidate \\
121291701801318272 & 54.440752 & 31.696937 & 3.23 & 3.71 & -6.09  & $\cdots$         & 2.78     & candidate \\
121352960919189248 & 53.088591 & 31.185584 & 2.98 & 3.80 & -6.58  & $\cdots$         & 1.32     & candidate \\
121389077299717120 & 52.539279 & 31.221953 & 3.54 & 6.68 & -9.74  & $\cdots$         & $\cdots$ & candidate \\
121394952814977280 & 52.550734 & 31.235416 & 3.43 & 7.20 & -11.12 & $\cdots$         & $\cdots$ & candidate \\
121396395923984256 & 52.523196 & 31.310432 & 3.40 & 6.61 & -9.56  & $\cdots$         & 0.26     & candidate \\
121399488300436352 & 52.485424 & 31.350081 & 3.49 & 6.47 & -10.23 & $\rm M4.1\pm0.4$ & 1.26     & member    \\
121404573540799232 & 52.212923 & 31.304325 & 3.56 & 7.68 & -9.23  & $\cdots$         & $\cdots$ & candidate \\
121412849943441024 & 52.358926 & 31.445404 & 3.18 & 7.52 & -9.92  & $\cdots$         & $\cdots$ & candidate \\
121465046681719552 & 53.397730 & 31.690906 & 3.56 & 6.87 & -8.76  & $\cdots$         & 0.37     & candidate \\
121485903042882304 & 53.553743 & 31.912094 & 3.20 & 3.34 & -6.05  & $\cdots$         & $\cdots$ & candidate \\
121528504822114176 & 53.089315 & 31.849377 & 3.30 & 4.86 & -5.42  & $\rm A3\pm0.2$   & 4.94     & member    \\
123883147629015424 & 50.570845 & 30.582361 & 2.82 & 6.73 & -7.64  & $\cdots$         & $\cdots$ & candidate \\
123910081367783552 & 50.786711 & 30.807329 & 3.74 & 7.63 & -7.81  & $\cdots$         & $\cdots$ & candidate \\
123959834268754176 & 50.553325 & 30.862014 & 3.36 & 6.80 & -7.71  & $\cdots$         & 1.00     & candidate \\
123967049813899008 & 50.852554 & 31.051460 & 3.36 & 7.91 & -7.57  & $\cdots$         & 1.98     & candidate \\
123984027820803456 & 50.724634 & 31.221461 & 2.95 & 7.95 & -7.86  & $\cdots$         & $\cdots$ & candidate \\
123984027820803584 & 50.725256 & 31.220559 & 2.97 & 8.68 & -8.65  & $\rm A1\pm0.2$   & 5.81     & member    \\
123990517515783680 & 51.393798 & 30.791119 & 3.28 & 7.84 & -8.80  & $\cdots$         & 7.10     & candidate \\
123991277725418752 & 51.294099 & 30.809658 & 3.45 & 8.03 & -9.09  & $\cdots$         & 1.30     & candidate \\
123993373669456000 & 51.335138 & 30.865771 & 3.30 & 8.35 & -8.69  & $\cdots$         & $\cdots$ & candidate \\
123993506812387712 & 51.360385 & 30.877058 & 3.45 & 8.37 & -8.40  & $\cdots$         & 0.84     & candidate \\
123994679339516800 & 51.281136 & 30.828328 & 3.52 & 8.14 & -8.31  & $\rm M2.8\pm0.3$ & 1.75     & member    \\
123995847570619520 & 51.229972 & 30.887357 & 3.25 & 7.59 & -7.96  & $\cdots$         & $\cdots$ & candidate \\
123995950649833216 & 51.281566 & 30.909474 & 3.42 & 8.15 & -7.80  & $\cdots$         & 1.00     & candidate \\
123996874066752640 & 51.379737 & 30.918846 & 3.54 & 8.16 & -8.33  & $\rm M5.2\pm0.2$ & 1.08     & member    \\
123998252752298752 & 51.302425 & 30.989481 & 3.41 & 8.37 & -8.21  & $\rm K9.5\pm1.6$ & 2.21     & member    \\
123999730221048320 & 51.458753 & 30.931711 & 3.43 & 8.33 & -8.29  & $\rm A1\pm0.2$   & 0.97     & member    \\
123999936379477504 & 51.445275 & 30.955693 & 3.44 & 8.24 & -8.06  & $\rm M4.6\pm0.6$ & 2.18     & member    \\
124002856957227648 & 51.582532 & 31.110303 & 3.40 & 8.18 & -8.24  & $\rm K7\pm1.6$   & 4.13     & member    \\
124009935063344128 & 51.245429 & 31.010398 & 3.44 & 8.43 & -7.98  & $\rm M3.1\pm0.7$ & 1.18     & member    \\
124016081160908416 & 51.156870 & 31.199361 & 3.12 & 6.78 & -8.66  & $\cdots$         & 3.06     & candidate \\
124017318111063808 & 51.277991 & 31.114673 & 3.40 & 7.89 & -7.92  & $\rm G7\pm0.5$   & 5.48     & member    \\
124017322407087360 & 51.277117 & 31.115575 & 3.31 & 7.10 & -8.20  & $\cdots$         & $\cdots$ & candidate \\
124017597284939776 & 51.262843 & 31.132227 & 3.40 & 7.74 & -8.18  & $\cdots$         & 2.24     & candidate \\
124018318839496064 & 51.407963 & 31.139139 & 3.33 & 7.81 & -8.33  & $\rm M0.2\pm0.9$ & 2.54     & member    \\
124018864299289088 & 51.457625 & 31.173290 & 3.52 & 7.98 & -8.16  & $\cdots$         & 0.66     & candidate \\
124030138589484672 & 51.617565 & 31.202155 & 3.42 & 8.21 & -8.02  & $\rm K7\pm1.5$   & 5.81     & member    \\
124030138589485440 & 51.615716 & 31.194618 & 3.45 & 7.95 & -8.04  & $\cdots$         & 0.96     & candidate \\
124034399197072896 & 51.930860 & 31.288977 & 3.41 & 8.34 & -10.24 & $\rm K4.5\pm0.6$ & 1.68     & member    \\
124034399197073024 & 51.926947 & 31.285487 & 3.42 & 8.28 & -10.42 & $\rm M2.8\pm0.3$ & 1.19     & member    \\
124475642661236736 & 52.371350 & 32.252125 & 3.60 & 7.12 & -9.40  & $\cdots$         & $\cdots$ & candidate \\
124475646957597952 & 52.373574 & 32.246963 & 3.62 & 7.26 & -9.43  & $\cdots$         & 0.62     & candidate \\
216418321101566464 & 56.521716 & 31.647741 & 3.18 & 4.50 & -5.91  & $\rm M0.3\pm0.8$ & 2.35     & member    \\
216497829534611072 & 55.924532 & 31.327956 & 2.74 & 6.25 & -7.89  & $\cdots$         & 0.60     & candidate \\
216508790291102464 & 55.727348 & 31.334727 & 3.36 & 6.36 & -8.57  & $\rm M3.1\pm0.3$ & 1.19     & member    \\
216514317913282304 & 55.787123 & 31.459427 & 2.83 & 5.05 & -6.86  & $\cdots$         & 2.44     & candidate \\
216520163364482560 & 55.972888 & 31.627645 & 3.45 & 6.39 & -9.52  & $\rm K5\pm0.3$   & 0.79     & member    \\
216523530618074624 & 55.857687 & 31.617337 & 3.32 & 4.96 & -5.64  & $\cdots$         & 3.47     & candidate \\
216524454036147072 & 55.880769 & 31.701164 & 3.06 & 4.75 & -7.35  & $\cdots$         & 5.97     & candidate \\
216527894305651840 & 55.328058 & 31.237570 & 3.41 & 6.47 & -8.50  & $\rm M2.3\pm0.4$ & 0.79     & member    \\
216543871583951744 & 55.018656 & 31.315787 & 3.42 & 6.28 & -8.42  & $\cdots$         & $\cdots$ & candidate \\
216563246180936704 & 55.613452 & 31.558551 & 3.80 & 6.77 & -8.96  & $\cdots$         & $\cdots$ & candidate \\
216572252727848320 & 55.733078 & 31.746196 & 3.12 & 5.00 & -6.93  & $\cdots$         & 4.51     & candidate \\
216572832547750784 & 55.786991 & 31.798894 & 2.82 & 5.15 & -7.54  & $\cdots$         & 2.69     & candidate \\
216572871202364160 & 55.807903 & 31.799214 & 3.20 & 7.13 & -9.74  & $\cdots$         & $\cdots$ & candidate \\
216575890564466304 & 55.781245 & 31.815758 & 3.46 & 4.54 & -6.29  & $\cdots$         & 0.18     & candidate \\
216585721745264896 & 55.360264 & 31.843630 & 3.13 & 4.09 & -6.06  & $\rm K7\pm0.4$   & 3.82     & member    \\
216587405372443904 & 55.493524 & 31.815683 & 3.34 & 4.74 & -6.96  & $\cdots$         & $\cdots$ & candidate \\
216587405372444032 & 55.494163 & 31.815292 & 3.30 & 4.04 & -6.54  & $\cdots$         & $\cdots$ & candidate \\
216588436163846144 & 55.642520 & 31.850269 & 3.00 & 5.73 & -7.57  & $\cdots$         & 3.42     & candidate \\
216588638027403904 & 55.659161 & 31.872866 & 3.18 & 3.95 & -6.01  & $\cdots$         & 5.90     & candidate \\
216588981624788480 & 55.685585 & 31.892166 & 3.53 & 5.49 & -7.06  & $\cdots$         & 7.25     & candidate \\
216590016712558080 & 55.588347 & 31.953305 & 3.39 & 5.70 & -7.76  & $\rm M2.5\pm1.1$ & 4.30     & member    \\
216590016712558208 & 55.586921 & 31.946872 & 3.40 & 4.84 & -8.34  & $\cdots$         & 8.11     & candidate \\
216592039641875712 & 55.377191 & 31.914551 & 3.29 & 2.51 & -6.37  & $\cdots$         & 10.37    & candidate \\
216601213692521472 & 56.191642 & 31.604388 & 3.25 & 5.13 & -6.06  & $\cdots$         & 4.12     & candidate \\
216606268869907072 & 56.495475 & 31.690701 & 2.84 & 4.63 & -5.69  & $\cdots$         & 2.49     & candidate \\
216608948929500032 & 56.393335 & 31.732071 & 3.03 & 4.24 & -5.92  & $\cdots$         & 4.20     & candidate \\
216613999810965760 & 56.014317 & 31.655093 & 3.21 & 4.02 & -6.58  & $\rm K7\pm0.4$   & 3.37     & member    \\
216614034170704256 & 56.001480 & 31.656658 & 3.67 & 3.54 & -7.39  & $\cdots$         & 3.91     & candidate \\
216617641943232128 & 56.027987 & 31.722923 & 3.22 & 4.33 & -5.22  & $\cdots$         & $\cdots$ & candidate \\
216621004901747584 & 56.108702 & 31.877509 & 3.66 & 4.97 & -6.43  & $\cdots$         & $\cdots$ & candidate \\
216636505439656960 & 56.498278 & 31.951167 & 2.99 & 3.23 & -6.25  & $\cdots$         & 2.65     & candidate \\
216637982908401792 & 56.555436 & 31.982448 & 2.82 & 4.55 & -5.32  & $\cdots$         & 2.33     & candidate \\
216643480466527744 & 56.693603 & 32.030345 & 3.16 & 4.73 & -5.39  & $\rm K7\pm1.2$   & 1.71     & member    \\
216647603635097856 & 56.760031 & 32.237265 & 3.17 & 4.37 & -5.74  & $\cdots$         & 1.78     & candidate \\
216649218542854400 & 56.347129 & 32.004377 & 3.12 & 6.60 & -10.07 & $\rm A5\pm0.2$   & 6.14     & member    \\
216659457744832512 & 56.762826 & 32.254421 & 3.27 & 4.31 & -5.79  & $\cdots$         & 1.87     & candidate \\
216661519329152000 & 56.566835 & 32.231462 & 3.07 & 3.88 & -5.40  & $\cdots$         & 2.03     & candidate \\
216662687560253696 & 56.502346 & 32.279487 & 3.00 & 3.90 & -5.56  & $\cdots$         & 2.54     & candidate \\
216662790639458816 & 56.543475 & 32.317686 & 3.23 & 4.58 & -6.52  & $\cdots$         & 2.29     & candidate \\
216662992501225984 & 56.487738 & 32.328021 & 3.17 & 4.13 & -6.53  & $\cdots$         & 2.07     & candidate \\
216667150030373376 & 55.924521 & 31.858398 & 3.70 & 4.38 & -5.90  & $\cdots$         & 1.75     & candidate \\
216668013318136320 & 56.005782 & 31.893081 & 2.92 & 4.67 & -6.26  & $\cdots$         & 6.16     & candidate \\
216668803593749888 & 55.961863 & 31.905859 & 3.27 & 4.86 & -6.11  & $\rm K3.6\pm0.7$ & 2.42     & member    \\
216670384140089088 & 55.841483 & 31.897773 & 3.29 & 4.09 & -5.58  & $\cdots$         & $\cdots$ & candidate \\
216673476518076288 & 55.862439 & 32.011624 & 3.37 & 5.58 & -6.30  & $\cdots$         & 1.97     & candidate \\
216674159417196416 & 56.072878 & 31.925330 & 3.00 & 4.20 & -6.92  & $\cdots$         & 5.81     & candidate \\
216683406482466688 & 55.732008 & 31.982894 & 3.22 & 3.89 & -6.62  & $\cdots$         & 1.43     & candidate \\
216685051453693440 & 55.732774 & 32.074545 & 3.69 & 4.27 & -6.74  & $\cdots$         & $\cdots$ & candidate \\
216687387916289280 & 55.651989 & 32.026197 & 3.12 & 4.61 & -6.15  & $\cdots$         & 4.46     & candidate \\
216687456635947648 & 55.632276 & 32.040875 & 2.85 & 4.04 & -6.46  & $\cdots$         & 2.45     & candidate \\
216688212549374208 & 55.663290 & 32.086597 & 3.41 & 4.03 & -6.32  & $\cdots$         & 2.60     & candidate \\
216689621299551488 & 55.707787 & 32.118134 & 3.27 & 4.79 & -5.79  & $\cdots$         & 1.89     & candidate \\
216697528334839936 & 55.736346 & 32.257784 & 3.55 & 3.83 & -6.43  & $\cdots$         & 1.10     & candidate \\
216698760989269120 & 55.901556 & 32.318429 & 3.08 & 4.48 & -5.30  & $\cdots$         & 1.80     & candidate \\
216703743151499136 & 56.393745 & 32.297036 & 3.31 & 3.94 & -6.42  & $\cdots$         & 0.14     & candidate \\
216708901408427904 & 56.256046 & 32.390130 & 3.22 & 4.62 & -6.44  & $\cdots$         & 1.00     & candidate \\
216709515586972800 & 56.454296 & 32.316288 & 3.33 & 4.61 & -5.12  & $\cdots$         & 1.90     & candidate \\
216710168422013440 & 56.450990 & 32.345797 & 3.28 & 3.67 & -6.70  & $\cdots$         & 0.04     & candidate \\
216710172718592640 & 56.447895 & 32.349759 & 3.12 & 3.98 & -5.41  & $\cdots$         & 2.65     & candidate \\
216710447596506880 & 56.383900 & 32.321349 & 3.31 & 3.67 & -5.27  & $\cdots$         & 3.61     & candidate \\
216711409669168000 & 56.432740 & 32.409746 & 3.28 & 4.42 & -5.32  & $\cdots$         & 2.29     & candidate \\
216712337382175488 & 56.588681 & 32.454909 & 3.14 & 3.80 & -5.72  & $\cdots$         & 2.15     & candidate \\
216712371741832576 & 56.537043 & 32.441361 & 3.17 & 4.86 & -7.33  & $\cdots$         & 2.70     & candidate \\
216712814121987712 & 56.502375 & 32.437318 & 3.59 & 4.11 & -6.02  & $\cdots$         & $\cdots$ & candidate \\
216714158448241792 & 56.347417 & 32.410264 & 3.13 & 4.23 & -5.56  & $\rm K5.6\pm1.4$ & 2.72     & member    \\
216714467685880320 & 56.413167 & 32.440045 & 3.11 & 4.63 & -5.62  & $\cdots$         & $\cdots$ & candidate \\
216714467686233344 & 56.411139 & 32.438565 & 3.30 & 4.50 & -5.50  & $\cdots$         & $\cdots$ & candidate \\
216716250095746816 & 56.420118 & 32.489109 & 3.69 & 3.49 & -6.71  & $\cdots$         & 2.63     & candidate \\
216716902930786432 & 56.466634 & 32.542254 & 3.14 & 3.87 & -5.82  & $\cdots$         & 1.92     & candidate \\
216718178536050176 & 56.079724 & 32.288246 & 3.07 & 6.09 & -9.73  & $\cdots$         & $\cdots$ & candidate \\
216721442712788480 & 56.123493 & 32.413043 & 3.20 & 5.23 & -5.52  & $\cdots$         & 2.47     & candidate \\
216721545790356864 & 56.124912 & 32.433028 & 3.16 & 3.44 & -5.58  & $\rm A1\pm0.2$   & 3.35     & member    \\
216721850733091968 & 56.112591 & 32.455298 & 2.77 & 3.84 & -7.37  & $\cdots$         & $\cdots$ & candidate \\
216723568719908992 & 56.019145 & 32.468395 & 3.39 & 4.44 & -5.86  & $\rm M3.3\pm0.9$ & 2.17     & member    \\
216724088412592640 & 55.946864 & 32.468929 & 3.53 & 6.76 & -9.78  & $\cdots$         & 0.39     & candidate \\
216724668231549056 & 55.972461 & 32.509377 & 3.11 & 3.68 & -6.54  & $\cdots$         & 2.20     & candidate \\
216724775605757312 & 56.085677 & 32.461030 & 3.24 & 5.00 & -6.59  & $\cdots$         & 2.17     & candidate \\
216725424145894400 & 56.142269 & 32.517465 & 2.97 & 4.23 & -6.49  & $\cdots$         & 1.80     & candidate \\
216729139294155520 & 56.340596 & 32.535158 & 3.37 & 4.24 & -6.63  & $\rm M3.6\pm0.3$ & 2.56     & member    \\
216729963927870848 & 56.355483 & 32.614519 & 3.48 & 4.27 & -5.83  & $\cdots$         & 3.56     & candidate \\
216731437100020096 & 56.161879 & 32.557035 & 3.13 & 4.04 & -5.75  & $\cdots$         & 1.65     & candidate \\
216734014080566400 & 56.238484 & 32.681098 & 2.79 & 3.25 & -6.06  & $\cdots$         & 2.65     & candidate \\
216829401012198912 & 57.012845 & 32.086724 & 3.49 & 6.24 & -9.91  & $\rm K7\pm1.1$   & 0.40     & member    \\
216831806193877504 & 57.093931 & 32.187523 & 3.58 & 6.63 & -9.71  & $\cdots$         & 0.81     & candidate \\
216837643052641024 & 57.280912 & 32.264721 & 3.56 & 6.88 & -10.88 & $\cdots$         & $\cdots$ & candidate \\
216837647349542272 & 57.280407 & 32.264269 & 3.54 & 6.32 & -9.45  & $\cdots$         & $\cdots$ & candidate \\
217048238186249088 & 56.882662 & 32.580982 & 3.79 & 7.04 & -9.38  & $\cdots$         & 0.61     & candidate \\
217049784373512448 & 56.890209 & 32.649548 & 3.54 & 7.28 & -9.82  & $\rm M4.3\pm0.3$ & 0.85     & member    \\
217068063754243584 & 57.453438 & 32.833972 & 2.94 & 3.69 & -5.37  & $\rm M2.8\pm0.4$ & 1.15     & member    \\
217072186920843264 & 57.213927 & 32.714366 & 3.75 & 6.23 & -9.09  & $\rm M2.2\pm0.4$ & 0.79     & member    \\
217088198561053440 & 56.629889 & 32.515255 & 3.37 & 3.57 & -6.80  & $\cdots$         & 2.94     & candidate \\
217090088346651136 & 56.674950 & 32.569818 & 3.32 & 4.30 & -5.27  & $\cdots$         & 3.05     & candidate \\
217090565086058624 & 56.761979 & 32.628750 & 2.72 & 3.67 & -5.12  & $\cdots$         & 5.03     & candidate \\
217095414105928192 & 56.584869 & 32.696801 & 3.40 & 6.02 & -10.08 & $\cdots$         & 0.55     & candidate \\
217104519434291200 & 56.746359 & 32.902363 & 3.59 & 4.44 & -6.12  & $\cdots$         & 1.72     & candidate \\
217112108642285184 & 56.396071 & 32.845285 & 3.43 & 2.94 & -7.37  & $\cdots$         & 2.58     & candidate \\
217130632837699584 & 56.947851 & 33.068172 & 3.67 & 4.68 & -7.31  & $\cdots$         & $\cdots$ & candidate \\
217144582891467776 & 56.850219 & 33.207827 & 3.75 & 6.14 & -9.73  & $\cdots$         & $\cdots$ & candidate \\
217164305382538240 & 57.561016 & 33.012804 & 3.71 & 6.75 & -9.97  & $\rm M3.7\pm0.2$ & 0.51     & member    \\
217165679772064384 & 57.629022 & 33.038139 & 3.49 & 2.89 & -6.91  & $\cdots$         & 0.90     & candidate \\
217304493112912256 & 54.705249 & 31.561875 & 3.18 & 3.93 & -6.33  & $\cdots$         & 3.85     & candidate \\
217307413689579392 & 54.830994 & 31.721273 & 3.53 & 5.27 & -6.90  & $\cdots$         & 5.61     & candidate \\
217313117407234944 & 54.911635 & 31.729285 & 3.28 & 4.60 & -6.41  & $\rm A5\pm0.2$   & 5.30     & member    \\
217317549813446656 & 54.650193 & 31.653468 & 3.49 & 7.18 & -8.87  & $\rm M1.8\pm0.6$ & 0.50     & member    \\
217325555632522624 & 54.829302 & 31.800202 & 3.22 & 3.03 & -6.44  & $\cdots$         & $\cdots$ & candidate \\
217325555632522880 & 54.827343 & 31.800138 & 3.21 & 3.67 & -7.13  & $\cdots$         & 1.74     & candidate \\
217328300115954560 & 54.967428 & 31.929547 & 3.27 & 5.10 & -6.78  & $\cdots$         & $\cdots$ & candidate \\
217328300116022528 & 54.967374 & 31.940529 & 2.93 & 5.53 & -6.11  & $\cdots$         & 5.54     & candidate \\
217336825626699008 & 55.342300 & 31.909405 & 3.28 & 3.83 & -6.47  & $\cdots$         & $\cdots$ & candidate \\
217344041171754368 & 55.308866 & 31.996147 & 3.31 & 4.58 & -6.08  & $\rm A5\pm0.1$   & 2.78     & member    \\
217346167179746176 & 55.423818 & 32.098545 & 3.11 & 4.28 & -5.72  & $\cdots$         & 0.01     & candidate \\
217346377633950464 & 55.412859 & 32.102204 & 3.19 & 4.52 & -5.46  & $\cdots$         & $\cdots$ & candidate \\
217347232331662976 & 55.324826 & 32.047493 & 3.32 & 3.99 & -6.28  & $\rm M3.5\pm0.4$ & 2.04     & member    \\
217348645376882816 & 55.229784 & 32.113858 & 3.16 & 4.50 & -6.51  & $\cdots$         & 1.49     & candidate \\
217349057693547008 & 55.253920 & 32.131725 & 3.27 & 4.42 & -6.26  & $\rm F0\pm0.1$   & 6.85     & member    \\
217352592451052800 & 54.981498 & 31.976198 & 3.68 & 5.05 & -5.63  & $\cdots$         & 4.35     & candidate \\
217358644060557696 & 55.065930 & 32.154617 & 3.21 & 4.71 & -6.97  & $\rm A1\pm0.2$   & 2.75     & member    \\
217358747139771264 & 55.087071 & 32.165668 & 3.11 & 5.01 & -7.14  & $\rm M1.7\pm0.6$ & 2.10     & member    \\
217363553207422592 & 55.244151 & 32.230972 & 2.92 & 5.13 & -6.11  & $\cdots$         & 1.74     & candidate \\
217363557503130240 & 55.231857 & 32.235148 & 3.20 & 4.27 & -5.80  & $\cdots$         & 2.11     & candidate \\
217383550574612864 & 54.554583 & 32.091862 & 3.56 & 4.89 & -7.50  & $\cdots$         & $\cdots$ & candidate \\
217407469248673152 & 54.955579 & 32.287333 & 3.13 & 4.90 & -7.10  & $\cdots$         & 1.86     & candidate \\
217415715585859328 & 55.099042 & 32.416592 & 3.10 & 3.91 & -5.66  & $\cdots$         & $\cdots$ & candidate \\
217440725180229376 & 55.515644 & 32.182854 & 3.31 & 4.18 & -6.31  & $\cdots$         & 5.33     & candidate \\
217440896978922112 & 55.475918 & 32.192990 & 3.28 & 4.69 & -7.20  & $\cdots$         & 3.38     & candidate \\
217444440327486848 & 55.492348 & 32.238318 & 3.22 & 4.57 & -7.43  & $\rm G6\pm0.4$   & 2.79     & member    \\
217446188378685824 & 55.574839 & 32.309479 & 2.88 & 4.04 & -6.44  & $\cdots$         & 1.02     & candidate \\
217455469803469056 & 55.765457 & 32.519854 & 2.84 & 4.61 & -6.53  & $\cdots$         & $\cdots$ & candidate \\
217457565747541120 & 55.411412 & 32.317215 & 3.32 & 4.17 & -6.25  & $\cdots$         & 1.76     & candidate \\
217458493460476928 & 55.310242 & 32.362894 & 3.18 & 4.31 & -5.93  & $\rm M1.1\pm0.9$ & 1.99     & member    \\
217463643124614656 & 55.382460 & 32.472199 & 3.26 & 4.53 & -6.14  & $\cdots$         & 3.11     & candidate \\
217465876507610112 & 55.536150 & 32.474356 & 3.33 & 4.21 & -5.72  & $\cdots$         & 2.86     & candidate \\
217466361840541952 & 55.587579 & 32.494643 & 3.10 & 4.14 & -6.36  & $\cdots$         & 1.70     & candidate \\
217475123573808000 & 55.895627 & 32.526988 & 3.22 & 5.03 & -5.74  & $\rm G5\pm0.5$   & 2.98     & member    \\
217479414244568320 & 55.810629 & 32.592766 & 3.43 & 3.57 & -6.02  & $\cdots$         & 2.42     & candidate \\
217479521620315264 & 55.806566 & 32.607210 & 3.11 & 3.08 & -3.37  & $\rm F0\pm0.2$   & 3.22     & member    \\
217481269670674048 & 55.887565 & 32.642586 & 2.89 & 4.39 & -5.82  & $\cdots$         & 3.26     & candidate \\
217482850218656512 & 55.893907 & 32.724620 & 3.32 & 4.11 & -7.29  & $\cdots$         & 1.28     & candidate \\
217510376665297152 & 55.243254 & 32.506634 & 3.43 & 6.63 & -9.60  & $\rm K5\pm0.4$   & 0.54     & member    \\
217510892061374208 & 55.156662 & 32.497486 & 3.40 & 7.17 & -9.08  & $\cdots$         & $\cdots$ & candidate \\
217863938374329472 & 56.342371 & 32.921457 & 3.25 & 7.45 & -10.73 & $\cdots$         & 0.76     & candidate \\
\bottomrule
\end{longtable*}

\bibliography{References}
\bibliographystyle{aasjournal}

\end{document}